\documentclass[preprint,5p,times,twocolumn]{elsarticle}
 
\usepackage{amssymb}
\usepackage{amsmath}
\usepackage{xcolor}
\usepackage{enumitem}
\usepackage{fancyvrb}
\usepackage{threeparttable}
\usepackage{hyperref}
\usepackage{ulem}
\usepackage{listings}

\newcounter{bla}

 \usepackage{geometry}
\newcommand{\Mpl}{M_{\text{pl}}}

\journal{Computer Physics Communications}

 \newcommand{\preprintnumbers}{%
  \begin{tabular}[b]{@{}r@{}}
    KEK-QUP-2026-0017\\
    KEK-TH-2878
  \end{tabular}}

\makeatletter
\def\ps@pprintTitle{%
  \def\@oddhead{\hfill\footnotesize\smash{\preprintnumbers}}%
  }
\makeatother

\begin{document}

\begin{frontmatter}

\title{Transient PULSE:\\ Code for the Transient Propagation of UltraLight (pseudo)Scalars to Earth}

\author[a,b,c]{Jason Arakawa\corref{cor1}}
\ead{arakawaj@udel.edu}
\author[a]{Charles Cheung}
\author[f]{Muhammad Hani Zaheer}
\author[b,c,d,e]{Volodymyr Takhistov}
\author[a]{Marianna S. Safronova}

\cortext[cor1] {Corresponding author}
\address[a]{Department of Physics and Astronomy, University of Delaware, Delaware 19716, USA}
\address[b]{International Center for Quantum-field Measurement Systems for Studies of the Universe and Particles (QUP, WPI),
High Energy Accelerator Research Organization (KEK), Oho 1-1, Tsukuba, Ibaraki 305-0801, Japan}
\address[c]{Theory Center, Institute of Particle and Nuclear Studies (IPNS), High Energy Accelerator Research Organization (KEK), Tsukuba 305-0801, Japan}
\address[f]{Department of Electrical \& Computer Engineering, University of Delaware, Delaware 19716, USA}
\address[d]{Kavli Institute for the Physics and Mathematics of the Universe (WPI), \mbox{The University of Tokyo Institutes for Advanced Study, The University of Tokyo, Kashiwa, Chiba 277-8583, Japan}}
\address[e]{Graduate University for Advanced Studies (SOKENDAI), 
1-1 Oho, Tsukuba, Ibaraki 305-0801, Japan}

\begin{abstract} 
We present Transient PULSE, a software package that numerically implements a previously developed source-to-detector framework for transient ultralight bosonic signals from astrophysical bursts. Predicting these signals relies on connecting the source emission spectrum to its evolution during propagation and the resulting observables at a terrestrial detector. The package discretizes a user supplied energy spectrum into momentum modes, propagates them through a specified matter density profile and constructs the time-frequency energy density distribution at Earth, accounting for dispersion and in-medium screening effects. This treatment accommodates non-Gaussian source spectra and provides inputs for detector specific analyses. A complementary analytical module estimates coupling sensitivities, arrival time delays  and screening thresholds from parameterized source and detector properties. 

\noindent {\bf PROGRAM SUMMARY}

\begin{small}
\noindent
{\em Program Title:} Transient PULSE \\
{\em CPC Library link to program files:} (to be added by Technical Editor) \\
{\em Developer's repository link:} \url{https://github.com/ccheung93/TransientPULSE} \\
{\em Code Ocean capsule:} (to be added by Technical Editor) \\
{\em Licensing provisions:} MIT License \\
{\em Programming language:} Python 3.8+ \\
{\em Supplementary material:} None \\
{\em Nature of problem:} Detecting a burst of ultralight bosonic particles emitted by a transient astrophysical event requires knowing how the emitted spectrum is reshaped by dispersive, in-medium propagation to a terrestrial detector, and what the resulting signal implies for the boson's couplings to the Standard Model. \\
{\em Solution method:} The source emission spectrum is discretized into momentum modes, each numerically propagated through a density profile to obtain its time of arrival at Earth and summed into a time-frequency energy-density distribution of the signal. A complementary analytical module computes detection-threshold, time-delay, and screening couplings as functions of frequency from parameterized source and experiment properties, producing sensitivity projections for the boson's couplings to the Standard Model particles. \\
{\em Unusual features:} includes environmental screening and time-dependent sources \\
{\em Restrictions:} The detection treatment is a simplified rescaling of DM sensitivities rather than a full experimental data-analysis pipeline. The achievable spectrogram time resolution is limited by available memory when the propagated signal duration greatly exceeds the intrinsic burst duration. \\
{\em Additional comments:} None 

\end{small}


\end{abstract}

    \end{frontmatter}

\begin{figure*}
    \centering
    \includegraphics[width=0.9\linewidth]{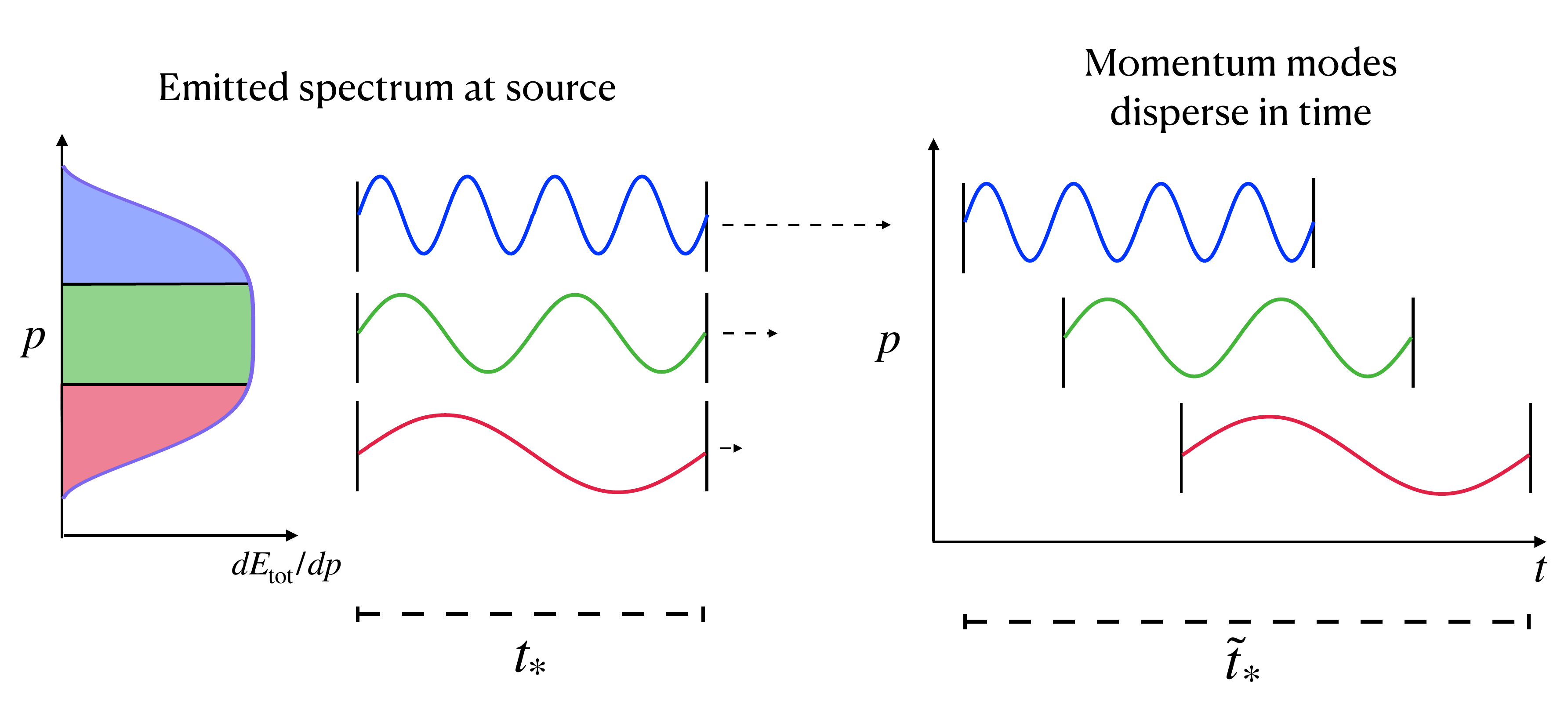}
    \caption{Schematic representation of transient signal propagation. A source emission spectrum as a function of momentum is discretized into momentum modes, each   numerically propagated from the source to the detector separately to obtain its arrival time and energy density contribution at Earth. The distinct velocity of each momentum mode modifies the original signal into a resulting ``anti-chirp" signal, where the highest frequency modes arrive first  and the frequency of the signal drifts to lower frequencies.}
    \label{fig:spectrum_prop_method}
\end{figure*}

\section{Introduction}

The quest for uncovering the nature of dark matter (DM) that comprises the dominant constituent of matter in the Universe as well as searches for hypothetical novel fields beyond the Standard Model have been at the forefront of particle physics, astrophysics and cosmology research. Searches for hypothetical weakly-coupled ultralight bosonic (ULB) fields with masses $\ll\mathcal{O}$(eV), which may contribute to DM abundance, have benefited from interdisciplinary efforts from precision sensors including atomic, molecular and optical experiments enabling sensitive searches for small effects of these fields. The experimental program on this front is thus rapidly developing. However, many of such detection schemes focus on searching for non-relativistic signals associated with the persistent local DM abundance near Earth's vicinity.

In contrast to searches for non-relativistic cold DM, astrophysical and new physics processes can produce energetic bursts of ULBs that propagate to terrestrial experiments, providing a distinct opportunity for their  detection~\cite{Dailey:2020sxa,Eby:2021ece,Arakawa:2023gyq,Arakawa:2024lqr,Arakawa:2025hcn,Khamis:2024oqa}. Searches for  bursts of exotic fields accompanying multimessenger events with quantum sensor networks were proposed in Ref.~\cite{Dailey:2020sxa}. A source-to-detector framework for relativistic  axion bursts from concrete transient astrophysical sources was introduced in Ref.~\cite{Eby:2021ece}, incorporating source emission, propagation and wave spreading, signal duration and coherence at the detector and experimental sensitivity. This program has been extended to scalar bosenova signals with different coupling structures~\cite{Arakawa:2023gyq,Arakawa:2024lqr}, and further developed for searches with quantum sensors, including in-medium propagation and screening effects~\cite{Stadnik:2021qyf}  in Ref.~\cite{Arakawa:2025hcn}. Accumulation of transient events from distinct sources can also result in persistent diffuse backgrounds~\cite{Eby:2024mhd}, different from cold DM.

Transient signals associated with physics beyond the Standard Model have several complementary features compared to typical signals associated with ambient cold DM. Transient signals can encode information about fundamental particle properties that is inaccessible or challenging to probe in typical searches for a persistent cold DM background~\cite{Eby:2021ece}. Further, the local DM density in the Solar System is not directly known~\cite{Gron:1995rn,Anderson:1995dw,Khriplovich:2006sq,Frere:2007pi,Pitjev:2013sfa,Tsai:2022jnv}. Transient signals originating elsewhere can provide a complementary search approach that does not rely on a persistent local ULB abundance.  More generally, a ULB need not constitute an appreciable component of DM. Production in energetic astrophysical events may enable probes of fields that would otherwise be difficult to detect locally due to suppressed abundance. 

The present work introduces TransientPULSE, a numerical implementation of the source-to-detector framework developed in Ref.~\cite{Eby:2021ece} and incorporating the in-medium propagation treatment of Ref.~\cite{Arakawa:2025hcn}. The package resolves a   source energy spectrum determined by the user into momentum modes, propagates them through a specified matter density profile and constructs the resulting time-frequency energy density distribution at the detector site. This enables direct numerical treatment of non-Gaussian source spectra and spatially varying propagation media within the existing framework. The package also implements the analytical estimates of coupling sensitivities, arrival time delays and screening thresholds developed in Refs.~\cite{Eby:2021ece,Arakawa:2025hcn}.

 This paper is organized as follows. In Sec.~\ref{sec:theory} we review the theoretical framework for the propagation of transient ULB signals from source to detector. In Sec.~\ref{sec:method} we describe the numerical implementation and the main output of the software package, the spectrogram of the energy density at the detector  and illustrate it with propagation examples. Sec.~\ref{sec:sensitivity} presents the projected sensitivity workflow and examples. Sec.~\ref{sec:codearchitecture} outlines the code architecture. In Sec.~\ref{sec:bosenova} we apply the full propagation pipeline to a physically motivated bosenova spectrum. Further details on particle physics properties and couplings, in-medium effects, source characterization, propagation  and detection are given in the Appendices. 

Due to the one-to-one correspondence between the propagating signal frequency   modes, momentum modes  and energy modes, we use them interchangeably in qualitative descriptions. We employ natural units  $c = \hbar = 1$ throughout. 

\section{Theoretical Background}
\label{sec:theory}

Energetic astrophysical events (see e.g.~\cite{Eby:2024mhd} for discussion), such as explosive emission from boson stars (bosenovae)~\cite{Eby:2016cnq,Levkov:2016rkk,Eby:2021ece,Arakawa:2023gyq,Arakawa:2024lqr}, emission from superradiant clouds around rotating black holes including in binary systems~\cite{Yoshino:2012kn,Baumann:2021fkf,Takahashi:2024fyq}, black hole formation from stellar collapse~\cite{deGiorgi:2024pjb,deGiorgi:2026lez}, and supernovae~\cite{Raffelt:1999tx}, can emit copious amounts of ULBs  in the form of  large amplitude  relativistic waves. Emitted   $\phi(t,x)$ waves can be characterized by an emission differential energy spectrum $dE/dp$ providing the distribution of frequencies, or equivalently energies and momenta, that are present. Focusing on transient signals we consider that the duration of the emission at the source lasts a finite time $t_*$.

The emission and propagation analysis we consider   applies equally to scalar ULBs and to pseudoscalar axion-like particles (ALPs). In both cases the propagation is governed by the dispersion relation, with the particle nature entering only through the coupling dependent in-medium contributions to the effective mass. However, the distinction is  important for detection since scalars and ALPs couple to different Standard Model operators and are probed by different experiments (see~\ref{app:particlephysics}).

As an emitted wave propagates to Earth, its shape changes due to composition of a distribution of frequency modes each traveling with a distinct propagation speed. The highest momentum modes arrive first and the lowest momentum modes lag behind, leading to wave spreading~\cite{Dailey:2020sxa,Eby:2021ece}. When wave spreading is significant, the signal at Earth takes the form of an ``anti-chirp''. Namely, the signal arrives first at high frequency and its frequency then decreases monotonically over the duration of the signal~\cite{Khamis:2024oqa}. This is illustrated schematically in Fig.~\ref{fig:spectrum_prop_method}. As a result of wave spreading, the duration of the signal at Earth  $\tilde{t}_*$  exceeds the intrinsic duration $t_*$ at the source.

The arrival time at Earth of each of the frequency (momentum) modes must be tracked in order to construct the time-frequency energy density signal distribution at the detector's location. The time   of arrival $t_{\rm \phi}$ for each momentum mode $p$ depends on the velocity of that mode $v_{\phi}$
\begin{align} \label{eq:t_arrival}
    t_{\rm \phi}(p) \simeq \int_0^{r} \frac{d\ell}{v_{\phi}(\ell,p)}  
\end{align}
where $\ell$ is the position along the line-of-sight from the Earth to the source, a distance $r$ away, and
\begin{align} \label{eq:group_vel}
    v_{\phi}(\ell) = \sqrt{1 - \frac{m^2 + \beta(\ell)}{\omega_p^2}}~,
\end{align}
where $\omega_p$ is the energy of the momentum mode  contained within the wavepacket $\phi$, and $m$ is the mass of $\phi$. 

We take into account in-medium effects that can decrease the velocity of the propagating waves via $\beta \propto \rho_{\rm matter}$, where $\rho_{\rm matter}$ represents the energy density of matter along the $\phi$ wave's trajectory, such as the density of hydrogen gas in galaxies (see \ref{app:particlephysics} for more details). 
If $\beta(\ell)$ becomes sufficiently large such that $m^2 + \beta(\ell) \geq \omega_p^2$ at any point along the trajectory, Eq.~\eqref{eq:group_vel} yields a vanishing or imaginary velocity  signaling that the mode cannot propagate through that region of the medium. Such momentum modes are  screened and excluded entirely from our calculation of the propagated signal, rather than merely slowed.
Here, we consider a stationary medium group velocity treatment and neglect potential effects such as partial reflection.

The signal amplitude plays a significant role for detection. While the code numerically tracks the propagated signal, the effects can also be estimated analytically. Consider the amplitude $\phi_0$ of a wavepacket $\phi$ given by
\begin{align}\label{eq:phi0}
    \phi_0 \simeq \frac{\sqrt{2\rho_{\phi}}}{\omega_{\phi}}.
\end{align}
Assuming spherically symmetric emission the energy density of the ULB field wavepacket $\phi$ is given by
\begin{align}
\label{eq:rho_approx}
    \rho_{\phi} \simeq \frac{E}{4\pi r^2 \tilde{x}_{*}}~,
\end{align}
where $E$ is the total energy emitted in the wavepacket, $r$ is the distance from the source to detector, and $\tilde{x}_{*}$ is the spatial thickness of the emitted spherical shell as it reaches terrestrial detector. 

This simplified approach does not capture the full details of a signal's propagation, whereas the numerical propagation workflow presented in Sec.~\ref{sec:method} calculates the momentum mode-resolved arrival time and energy density information, providing an accurate representation of the signal characteristics relevant for terrestrial detectors. However, Eqs.~\eqref{eq:phi0} and \eqref{eq:rho_approx} enable straightforward estimates and interpretation of projected sensitivities. This is used in the 
projected sensitivity plotting workflow (see Sec.~\ref{sec:sensitivity}) to   estimate detection sensitivities, as detailed in \ref{app:detection}. The main output of the propagation workflow is the \textit{spectrogram}, the energy density of the propagated signal in the frequency-time plane, considering the dispersion and in-medium effects discussed above. 

For a given emission spectrum $dE/dp$, the time-dependent density for a given momentum mode bin $p_i$ of width $\Delta p$ is 
\begin{align}
\label{eq:rho_pt}
    \rho_{\phi}(p_i,t) = \frac{1}{4\pi r^2 t_*\sqrt{1-m^2/\omega_i^2}}  \frac{dE}{dp}(p_i) \Delta p~\Theta(t,t_{\phi}(p_i))~.
\end{align}
Here, the density normalization  uses the vacuum group velocity at the observation point. 
The  $\Theta$ function produces an anti-chirp frequency drift, since it takes into account the differing time of arrivals for each $p_i$. That is,
\begin{align}
    \Theta(t,t_p) = \theta(t - t_p) - \theta(t-\{t_p + t_*\})
\end{align}
where $\theta(t)$ is the Heaviside step function. Eq.~\eqref{eq:rho_pt} is a generalization of Eq.~\eqref{eq:rho_approx}. One can then calculate the momentum summed density as a function of time as
\begin{align}
    \rho_{\phi}(t) = \sum_{i} \rho_{\phi}(p_i,t)~.
\end{align}
The normalization of $\rho_{\phi}(p_i,t)$ already includes $\Delta p$, which implicitly keeps track of the number of momentum bins $N_p$. Below we display $\rho(t)$ considering several examples, which can be integrated over time directly for a given experiment with an appropriate sensitivity. While this is beyond the scope of the functionality of the code the approach can be used for signal analyses inputs.

The  time-of-arrival calculations have implications for multimessenger astronomy observations, where it is of interest to find the difference in arrival times between known signals, such as gravitational waves with propagation speed $v_{\rm GW} = 1$, and potentially coincident $\phi$ signals. In general, Eq.~\eqref{eq:t_arrival} is evaluated numerically.

\section{Methodology}
\label{sec:method}

The workflow begins with defining a source emission spectrum. Users instantiate a \texttt{SpectrumSource} object, typically a \texttt{CSVSpectrum} loaded from simulations or other data, an \texttt{AnalyticSpectrum} for idealized studies, or a \texttt{TimeVaryingSpectrum} for time-dependent emissions\footnote{See Sec.~\ref{sec:tdgaussian} for brief discussion about the scope of the time-dependent emissions}, which provides arrays of momenta $p_i$ and corresponding amplitudes $A_i = (dE/dp)_i$ that characterize the source emission. Next, users create configuration objects: a \texttt{PhysicsConfig} instance specifies the ULB field mass, coupling strength, energy density fraction $K$, and burst duration, while a \texttt{PropagationConfig} instance points to a density profile CSV file and sets the interpolation resolution. This separates the physical model of the source spectrum from the details of the numerical implementation.

With the spectrum and configurations defined, users instantiate a \texttt{WaveformCollection} object that bundles all inputs. Calling the \texttt{propagate\_all()} method on this object executes the full propagation pipeline, using vectorized NumPy operations for efficiency (implemented in \texttt{waveform/propagation.py}). The method internally loads and interpolates the  matter density profile $\rho_{\rm matter}(r)$ from a CSV file onto a uniform grid spanning the distance between source and detector. For each mode with momentum $p_i$ and mass $m$, the code computes the energy $\omega_i=\sqrt{p_i^2+m^2}$ and evaluates the group velocity of Eq.~\eqref{eq:group_vel} at each position\footnote{Note that although the group velocity (and therefore the momentum) of $\phi$ changes as a function of position, $\omega_i$ is fixed throughout the propagation, and therefore we calculate the energy at the emission with the momentum modes defined by the source spectrum.}, obtaining the arrival time for each mode by numerically integrating the inverse group velocity along the path (see Eq.~\eqref{eq:t_arrival}). The arrival time for each frequency mode is used to construct the spectrogram, discussed below. The \texttt{propagate\_all()} method returns a dictionary containing the time range, energy (frequency) array,   and spectrogram data. This data structure can be passed directly to the plotting functions.  

To build the spectrogram, the density of the field at  the location of a terrestrial detector is calculated and filled into the spectrogram over the duration $t_*$ for each momentum mode according to Eq.~\eqref{eq:rho_pt}. All bins are filled with the same constant value for a given mode.  For temporally resolved emission windows the spectrogram records the density given by Eq.~\eqref{eq:rho_pt}. Unresolved windows receive the coarse bin correction described in Sec.~\ref{ssec:coarse}.

A key quantity for analysis is the energy density of the field as a function of time, summed over momenta (frequencies). To construct it the energy in each momentum bin is spread uniformly over its emission window $t_\ast$, starting at the arrival time of that mode, so that each mode contributes a constant density for the duration of its window. The number of momentum bins enters only through the bin width $\Delta p$ in the normalization of Eq.~\eqref{eq:rho_pt}. In addition to the two-dimensional spectrogram we display its two projections. The frequency-summed projection gives the energy density as a function of time $\rho_\phi(t)$. The time-summed projection gives the time-integrated energy density in each momentum bin, this is the source spectrum weighted by the inverse group velocity, with screened modes removed. We normalize it to a unit peak value, since we use it only to determine the mean and standard deviation of the frequencies present in the signal. The spectrogram contains no phase information, since the input spectrum we consider does not specify one.

The resulting spectrogram plot displays frequency versus time with color indicating density and therefore signal strength, revealing how different spectral components arrive at different times due to dispersion through the medium defined by the density profile. Additionally, computational parameters are recorded in a log file, along with timing information and any warnings or errors encountered during the calculation.

The workflow can export   to a file the source parameters (using \texttt{export\_source\_parameters()}) that encode  the effective astrophysical parameters implied by the propagation computation. This bridges to the projected sensitivity plotting workflow discussed in the next section.

\subsection{Time-Independent Gaussian Signal Example}
\label{sec:example1}

The notebook \texttt{examples\_propagation.ipynb} provides three executable examples, including the time-independent and time-dependent Gaussian signal sources discussed in this section. In each case, the user sets the physical parameters at the top of the relevant cell, runs all cells in order, and the notebook saves the spectrogram plot and a \texttt{.param} file for downstream use with the projected sensitivity plotting workflow.

\begin{figure*}
    \centering 
    \includegraphics[width = 0.9\linewidth]{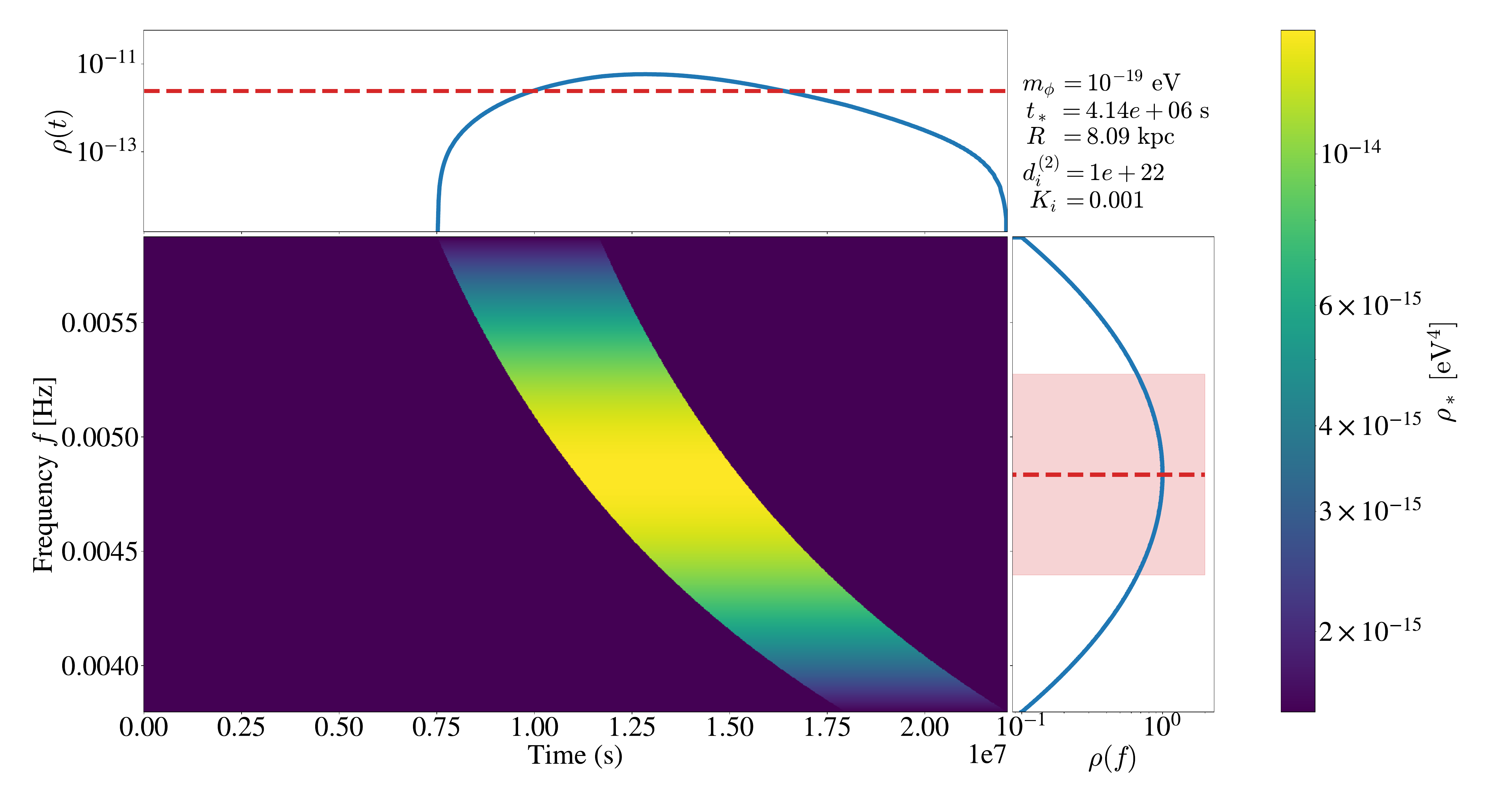}
    \caption{Spectrogram of a time-independent Gaussian signal wavepacket emission spectrum at Earth. This displays the frequency versus time dependence of the signal as it reaches the detector. The color bar represents the energy density of the field in each pixel in the time-frequency plane. The input parameters used here are displayed in the top right. The top panel represents the frequency-integrated energy density as a function of time, with the time-averaged energy density given by the red dashed line. The right panel shows the time-integrated frequency distribution, which is used to calculate the average frequency (red dashed) and standard deviation of frequencies (light red box).}
    \label{fig:time_independent_gaussian_spectrogram}
\end{figure*}

The Gaussian signal example models a burst with a constant in time Gaussian momentum distribution over the duration of the emission. The configuration parameters supplied by the user include:
\begin{itemize}
    \item \texttt{mass} -- ULB field mass $m$ in units of eV.
    \item \texttt{N\_periods} -- burst duration expressed in number of oscillation periods of the field. The notebook converts this to \texttt{burst\_duration} in seconds via $t_* = 2\pi N_{\rm periods}/(m\cdot\texttt{SEC\_TO\_INEV})$.
    \item \texttt{Etot} -- total emitted energy (e.g.\ in solar masses, converted to eV via \texttt{SOLAR\_TO\_EV}).
    \item \texttt{peak\_momentum} -- central momentum of the Gaussian spectrum. 
    \item \texttt{delta\_p} -- Gaussian spectrum momentum width, from which the peak amplitude is derived together with \texttt{Etot} to fix the total energy.
    \item \texttt{num\_points} -- number of momentum points used to sample the Gaussian spectrum.
    \item \texttt{coupling}, \texttt{K} -- dilatonic coupling strength and energy density fraction.
    \item \texttt{density\_num\_points} and \texttt{N\_points\_spectrogram} -- resolution parameters for the density profile interpolation and spectrogram time grid, respectively.
\end{itemize}

\subsection{Time-Dependent Gaussian Signal Example}
\label{sec:tdgaussian}

The time-varying example encapsulates a Gaussian spectrum in a \texttt{TimeVaryingSpectrum} routine that modulates the amplitude over a sequence of time windows. In addition to the Gaussian parameters above, the configuration parameters supplied by the user include:
\begin{itemize}
    \item \texttt{N\_time\_steps} -- number of time windows into which the burst duration is divided.
    \item \texttt{amplitude\_profile} -- a list of \texttt{N\_time\_steps} scale factors (one per window) that define how the overall emission amplitude varies in time. Any shape can be used; the example uses a Gaussian envelope centered at the midpoint.
    \item \texttt{time\_windows} -- a list of \texttt{(t\_start, t\_end)} pairs (in natural units) defining each window boundary. These are computed automatically from \texttt{burst\_duration} and \texttt{N\_time\_steps} in the notebook.
\end{itemize}
All other configuration parameters (\texttt{coupling}, \texttt{K}, \texttt{density\_num\_points}, \texttt{N\_points\_spectrogram}) are set the same way as in the time-independent scenario.  The time-dependent implementation considered here divides the burst into equal duration emission windows and varies the overall spectral amplitude between windows. The momentum dependence of the source spectrum remains fixed.  

\begin{figure*}[t]
    \centering 
    \includegraphics[width = 0.9\linewidth]{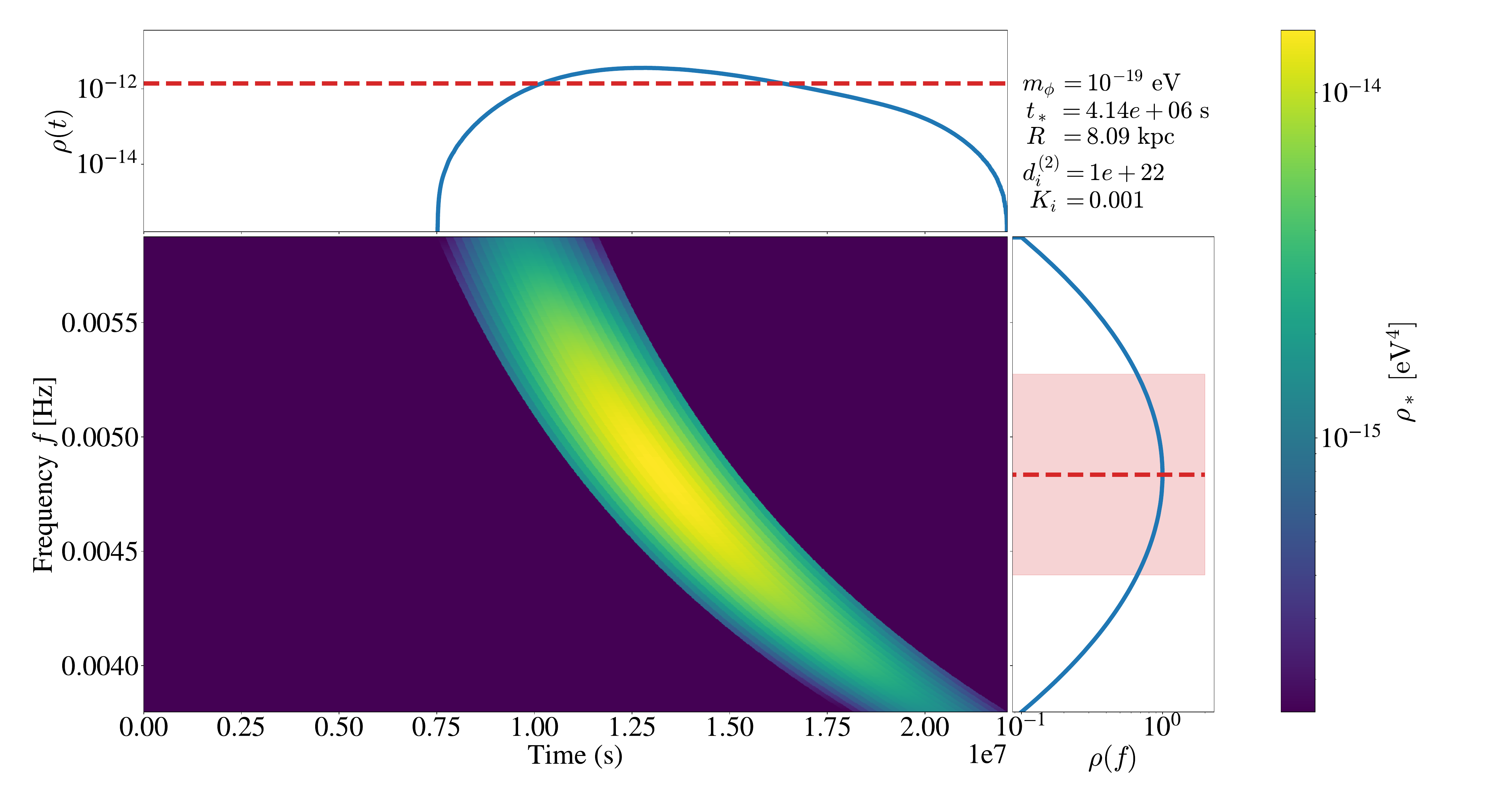}
\caption{Spectrogram of a time-dependent Gaussian signal wavepacket emission spectrum at Earth. This displays the frequency versus time dependence of the signal as it reaches the detector. The color bar represents the energy density of the field in each pixel in the time-frequency plane. The input parameters used here are displayed in the top right. The top panel represents the frequency-integrated energy density as a function of time, with the time-averaged energy density given by the red dashed line. The right panel shows the time-integrated frequency distribution, which is used to calculate the average frequency (red dashed) and standard deviation of frequencies (light red box).}
    \label{fig:time_dependent_gaussian_spectrogram}
\end{figure*}

\subsection{Gaussian Spectra Results}

We now present the propagation results for the time-independent and time-dependent Gaussian signal examples described above.
Fig.~\ref{fig:time_independent_gaussian_spectrogram} displays the time-independent Gaussian  emission spectrum, propagated over $r \simeq8.09$ kpc with benchmark values of $t_* = 4.14\times 10^{6}$ s and $m = 10^{-19}$ eV, with other parameters indicated in the figure.  Considering a possible coincidence emission with gravitational waves, the arrival time of $\simeq 7.5\times 10^{6}$~s is measured relative to the arrival of a gravitational wave emitted at the beginning of the burst.  The total duration of the signal becomes $\tilde{t}_{*} \simeq 1.5 \times 10^7$ seconds.   The upper panel shows the energy density summed over momenta at the detector, while the right panel shows the time-integrated contribution of each spectral bin, normalized to unit maximum. These benchmark parameters illustrate the propagation approach and can be varied by the user.

The propagated time-frequency spectrogram distribution can be used as input to time-frequency or excess power analyses in an experiment. Construction of a coherent matched filtering templates for analyses, such as those analogous to gravitational wave searches by LIGO~\cite{LIGOScientific:2016aoc, Romano:2016dpx, Panelli_2020}, requires additional phase and detector response information, which are beyond the scope of present work.

Similarly to Fig.~\ref{fig:time_independent_gaussian_spectrogram}, Fig.~\ref{fig:time_dependent_gaussian_spectrogram} displays the spectrogram for a time-dependent emission spectrum. Here, the spectrum was simulated with a Gaussian signal shape in momentum and time. All other parameters are the same as in the time-independent case.

\begin{figure*}[t]
    \centering
    \includegraphics[width = 0.9\linewidth]{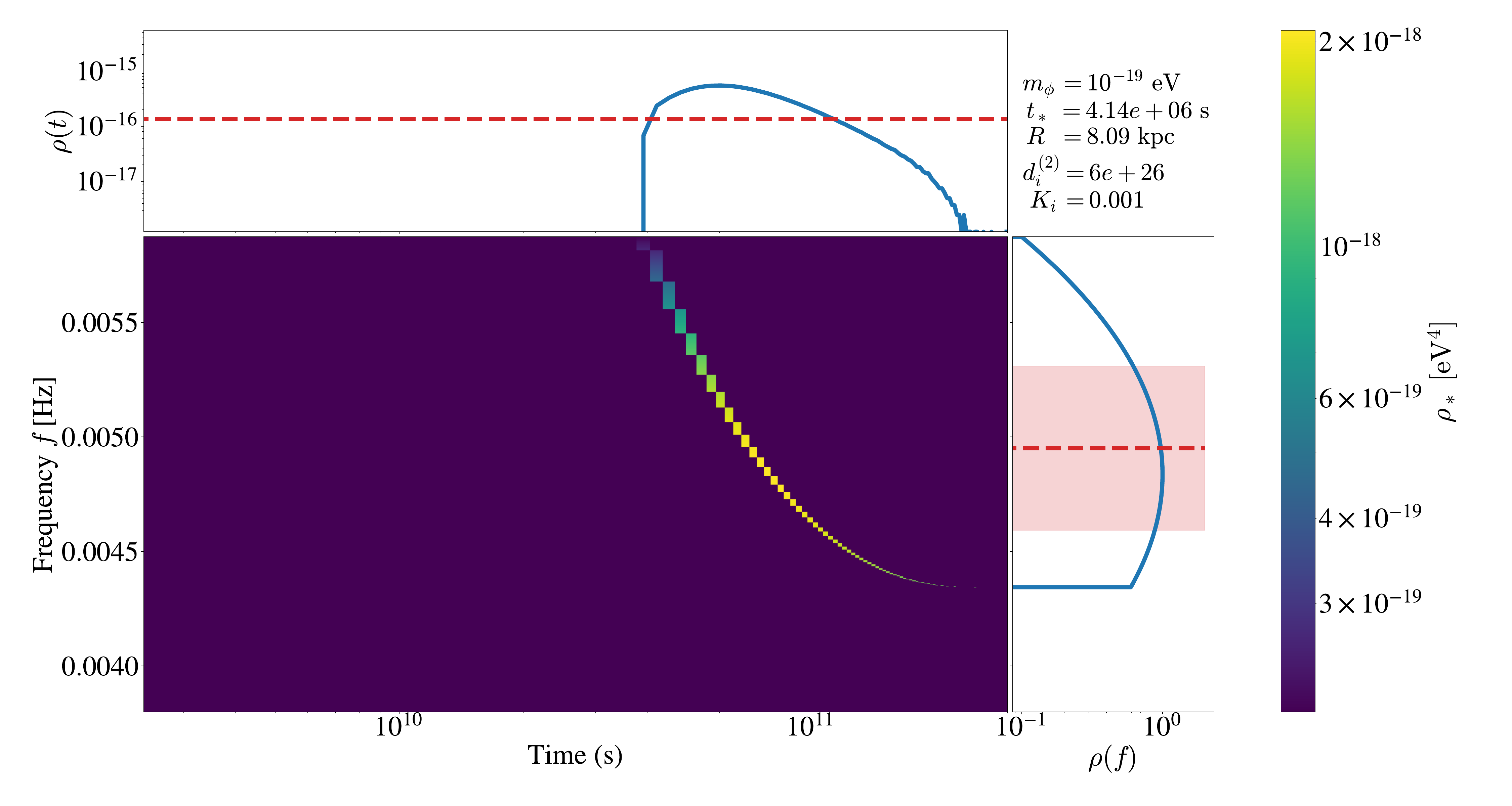}
    \caption{Spectrogram of a time-independent Gaussian wavepacket emission spectrum at Earth similar to Fig.~\ref{fig:time_independent_gaussian_spectrogram}, however in a scenario where part of the $\phi$ signal is screened. This displays the frequency versus time dependence of the signal as it reaches the detector. The color bar represents the energy density of the field in each pixel in the time-frequency plane. The input parameters used here are displayed in the top right. The top panel represents the frequency-integrated energy density as a function of time, with the time-averaged energy density given by the red dashed line. The right panel shows the time-integrated frequency distribution, which is used to calculate the average frequency (red dashed) and standard deviation of frequencies (light red box).}
    \label{fig:time_independent_gaussian_spectrogram_screened}
\end{figure*}

In Fig.~\ref{fig:time_independent_gaussian_spectrogram_screened}  we illustrate a scenario where the screening is sufficiently large such that some components of the signal spectrum do not propagate to the detector. A momentum mode is excluded from the propagated signal if $m^2+\beta \geq \omega_p^2$ anywhere along the trajectory in some region of the medium through which $\phi$ propagates. Note that for the same considered parameters, except for the increased coupling strength, the arrival times increases due to screening effects and the larger in-medium effective mass slows the propagating modes.

\subsection{Runtime and Memory Limitations} \label{ssec:coarse}

For variety of parameters considered the code execution runtime is of the order of a few seconds to minutes, sensitively depending on the number of computational grid points required. Storing the spectrogram requires memory of order $N_{t} \times N_p$, where $N_t$ and $N_p$ are the number  of time sample points and momentum bins,  respectively. For grid points $N_t \times N_p = 10^3\times 10^3$ the propagation calculation completes in $0.03$ seconds, and saving the spectrogram takes $1.84$ seconds. If we increase $N_t = 10^4$, the corresponding numbers are $0.05$ and $8.86$ seconds, while increasing $N_p = 10^{4}$ gives $0.27$ and $8.69$ seconds. The scaling with the number of points is roughly linear in both $N_t$ and $N_p$.

  Resolving the temporal structure requires a time spacing smaller than the emission window duration. This corresponds to having at least one grid point per $t_*$ assuming a static spectrum.  Then to keep track of the long duration signal, the time axis requires $N_t \geq \tilde{t}_*/t_*$. When the signal duration at terrestrial detector's location is much longer than the intrinsic duration of the burst at the source, i.e. $\tilde{t}_{*} \gg t_*$, the achievable resolution for the spectrogram's grid spacing, based on the memory of the computer, may not be sufficient. 

If the grid spacing is not sufficient to have at least one time bin per $t_*$, its contribution is assigned to a time bin with a density rescaled by the ratio of the window duration to the bin width, such that the energy per mode is properly conserved when changing the grid spacing.  This prescription retains the integrated contribution of an unresolved window, but does not resolve its temporal structure. For $t_* \gtrsim \Delta t$, the emission window is filled at the time grid points it contains. This is accurate when $t_* \gg \Delta t$  or when the arrival times of different modes are spread over many time bins so that discretization errors average out, as in the examples presented here.

If the memory is not sufficient, parameters and inputs can be adjusted. For instance, the regime $\tilde{t}_{*} \gg t_*$ can arise when the input spectrum includes a large range of momentum values, since this will affect the amount of wave spreading. The user can choose to limit the momentum domain, for example around any dominant peaks in the spectrum, such that the duration of the subset of the signal becomes significantly smaller. 

\section{Projected Signal Sensitivity}
\label{sec:sensitivity}

The projected sensitivity workflow determines which regions of parameter space, in coupling strength versus ULB energy, are accessible to detection for given source properties and detector capabilities, taking into account existing experimental limits. It is based on the analytical estimates of Refs.~\cite{Eby:2021ece,Arakawa:2025hcn} and allows rapid scans of parameter space without a numerical propagation calculation. The workflow can be used on its own, with source parameters specified directly, or with parameters derived from the propagation workflow of Sec.~\ref{sec:method} through the bridge functions described in Sec.~\ref{sec:codearchitecture}.

The workflow begins by defining the source parameters. Users create a \texttt{Source} object by specifying the total emitted energy (in units of solar masses $M_\odot$), ULB field mass, burst duration, distance to source, field type (scalar or ALP, which affects screening and allowed couplings), coupling type, and coupling order. These parameters define the simplified astrophysical scenario considered for the analytical procedure.

Next, users define the experimental configuration through an \texttt{Experiment} object, specifying the integration time for transient searches, integration time for DM searches, fractional frequency sensitivity, and measurable time delay thresholds (dictionary mapping time delay labels to threshold values in seconds). Optionally, users may also override the energy density fraction $K$ and the average interaction length per galaxy $D_g$ (see \ref{app:prop_analytical}), and consider a secondary $D_g$ value to shade a systematic band between the two screening assumptions. These parameters encode the detector capabilities.

Combining the source and experiment in a \texttt{SignalModel} object enables executing projected sensitivity calculations. The \texttt{SignalModel} constructor calls analytical formulas to compute the detection threshold coupling as a function of frequency (following the rescaling procedure detailed in~\ref{app:detection}), the time delay couplings for each time delay threshold, critical screening couplings for quadratic couplings (evaluating screening effects stemming from signal propagation through Earth, the atmosphere, and the apparatus environment), the energy-time uncertainty constraint $E_\textrm{unc}=2\pi/t_*$, and relevant astrophysical constraints for the specified coupling type.

Subsequently, users create a \texttt{Plot} configuration object specifying axis limits, styling options (legend configurations, label positions), and whether to exclude a mass region from the plot. An \texttt{OutputHandler} instance combines the plot configuration with the source-spectrum pair(s) and generates the sensitivity projection figure. The handler coordinates placement of all projected sensitivity plot elements: the detection threshold curve (black solid line), time delay curves (colored dashed lines), screening regions for quadratic couplings (blue shaded), astrophysical exclusions (gray shaded), uncertainty principle exclusion (brown shaded), experimental limits, and the viable parameter space (green shaded region bounded by detection threshold and time delay curves).

The \texttt{Source}/\texttt{Experiment}/\texttt{SignalModel}/\texttt{OutputHandler} classes described above are the recommended object-oriented entry point for the figures. The repository also includes \texttt{examples\_projected\_sensitivity\_plots.py}, a standalone Python script (independent of \texttt{SignalModel} and \texttt{OutputHandler}) that generates the benchmark $2\times2$ dilatonic coupling grids included in Ref.~\cite{Arakawa:2025hcn}.

\subsection{Grid Figures}
Instead of a single \texttt{Source} and \texttt{SignalModel} object, users can provide two-dimensional lists (nested lists of lists) representing a grid of parameters. For example, a $2\times2$ grid might vary mass along columns and burst durations along rows. The \texttt{OutputHandler} detects the grid structure and creates a figure with corresponding subplot layout.

Each panel shows the full projected sensitivity and constraint structure for its specific parameter combination. Mass values are displayed as column titles and burst durations as row labels on the right side. A shared legend appears once per figure rather than repeated in each panel. This layout enables an immediate visual comparison of how viable regions shift with parameter variations, relevant for evaluating the detection prospects across different astrophysical scenarios.

The grid system supports arbitrary rectangular layouts, automatically adjusting the figure size and subplot spacing for optimal readability. All styling options (colors, line styles, and label positions) apply consistently across panels.

\subsection{Examples of Configurations}
\label{sec:example2}

Generation of projected sensitivity figures by manually defining source parameters, without running waveform propagation, is demonstrated in notebook
\texttt{examples\_projected\_sensitivity\_plots\_manual.ipynb}. This includes examples below. The alternative route, deriving the \texttt{Source} parameters from a waveform propagation run instead of setting them manually, is demonstrated in \texttt{examples\_complete\_workflow.ipynb} (see Sec.~\ref{sec:method}).

\subsubsection{Example A: Single Emission Source}

The user instantiates a \texttt{Source} object with the following parameters:
\begin{itemize}
    \item \texttt{Etot} -- total emitted energy in units of solar masses.
    \item \texttt{mass} -- ULB field mass in eV.
    \item \texttt{tstar} -- burst duration in seconds.
    \item \texttt{R} -- source-to-Earth distance in parsecs.
    \item \texttt{ULB\_type} -- field type: \texttt{`scalar'} or \texttt{`ALP'}.
    \item \texttt{coupling\_type} -- Standard Model coupling channel: \texttt{`photon'}, \texttt{`electron'}, or \texttt{`gluon'} for scalar ULBs; \texttt{`photon'}, \texttt{`electron'}, \texttt{`proton'}, \texttt{`neutron'}, or \texttt{`EDM'} for ALPs.
    \item \texttt{coupling\_order} -- coupling order for scalar ULBs: \texttt{`linear'} or \texttt{`quad'}. Ignored for ALPs.
\end{itemize}

Next, the user creates an \texttt{Experiment} object specifying \texttt{integration\_time} (for transient searches), \texttt{integration\_time\_DM} (for DM searches), \texttt{sensitivity} (fractional frequency sensitivity), and \texttt{time\_delays} (a dictionary mapping labels to threshold values in seconds). The \texttt{Source} and \texttt{Experiment} are combined into a \texttt{SignalModel}, which computes all sensitivity curves.

Finally, a \texttt{Plot} object sets the axis limits (\texttt{xlims}, \texttt{ylims}) and display options (\texttt{exclude\_mass}, \texttt{include\_legend}). An \texttt{OutputHandler} is passed this configuration and calls \texttt{plot\_parameter\_space()} with the source, signal model, plot, and output path to save the figure.

\subsubsection{Example B: Multiple Source  $2\times 2$  Parameter Grid}

To compare sensitivities and constraints across parameter combinations, the user defines two lists -- \texttt{masses} and \texttt{times} -- and builds nested lists of \texttt{Source} objects, one per grid point, sharing a single \texttt{Experiment}. This example also illustrates the optional \texttt{Experiment} parameters: an explicit energy density fraction \texttt{K}, the interaction length per galaxy \texttt{Dg}, and a secondary \texttt{Dg\_secondary}, which causes a shaded uncertainty band to be drawn between the time-delay coupling curves computed at the two screening lengths. Nested lists of the resulting \texttt{SignalModel} objects are built in the same way and passed, together with the \texttt{Source} grid, directly to \texttt{plot\_parameter\_space()}, which produces a single figure with one subplot per combination. An optional \texttt{legend\_config} dictionary and a \texttt{label\_positions} dictionary (overriding the automatic placement of exclusion-region and screening-curve labels) can be supplied to the \texttt{Plot} object to fine-tune the grid's shared legend and per-panel annotations.

\section{Code Architecture}
\label{sec:codearchitecture}

The codebase implements the two complementary computational workflows described above: the detailed numerical propagation of Sec.~\ref{sec:method}, and the analytical projected sensitivity  of Sec.~\ref{sec:sensitivity}. We now describe how the code implementing both workflows is organized.

\subsection{System Requirements}
The code requires Python 3.8 or later version with the following dependencies:
\begin{itemize}[noitemsep]
    \item NumPy ($\geq 1.20$)
    \item Matplotlib ($\geq 3.3$)
    \item SciPy ($\geq 1.6$) 
    \item Jupyter (optional)
\end{itemize}

Development and testing were performed on Linux and macOS systems. Windows users may need to adjust path separators in configuration files. 

\subsection{Installation}
The code is distributed as a Python code repository via GitHub at \url{https://github.com/ccheung93/TransientPULSE}. Installation proceeds as follows:

\begin{lstlisting}[language=bash, caption={Installation commands}]
# Clone repository
git clone https://github.com/ccheung93/TransientPULSE.git
cd TransientPULSE

# Create virtual environment (recommended)
python3 -m venv .venv
source .venv/bin/activate

# Install dependencies
pip install -r requirements.txt
\end{lstlisting}

The virtual environment isolates package versions to ensure reproducibility. The \texttt{requirements.txt} file specifies the tested version ranges for each dependency.

\subsection{Package Structure}
The codebase is organized into a modular directory structure:

\begin{itemize}[noitemsep]
\item \texttt{calculations/} -- Analytical sensitivity formulas
  \begin{itemize}
  \item \texttt{sensitivities.py}
  \end{itemize}
\item \texttt{plotting/} -- Visualization and output management
  \begin{itemize}
  \item \texttt{plots.py}
  \item \texttt{limits.py}
  \item \texttt{output\_handler.py}
  \end{itemize}
\item \texttt{utils/} -- Shared utilities
  \begin{itemize}
  \item \texttt{constants.py} -- Physical constants and unit conversions
  \item \texttt{data\_utils.py} -- CSV loading and interpolation
  \item \texttt{expt\_params.py} -- Experimental benchmark values and projected sensitivity file paths
  \item \texttt{validation.py} -- Shared input validation helpers
  \item \texttt{logging\_utils.py} -- Logging configuration
  \end{itemize}
\item \texttt{inputs/} -- Input specifications and configurations
  \begin{itemize}
  \item \texttt{spectrum\_sources.py} -- Spectrum generation classes
  \item \texttt{configs.py} -- Physics and propagation configurations
  \item \texttt{source.py}, \texttt{experiment.py}, \texttt{spectrum.py} -- Projected sensitivity plotting inputs
  \end{itemize}
\item \texttt{waveform/} -- Numerical propagation
  \begin{itemize}
  \item \texttt{collection.py} -- Orchestrates propagation workflow
  \item \texttt{propagation.py} -- Core numerical integration, spectrogram generation and plotting
  \end{itemize}
\item \texttt{data/} -- Data files that include experimental constraints, matter density profiles, and emission spectra. See Sec.~\ref{sec:datafiles} for further details.
\item \texttt{examples\_projected\_sensitivity\_plots.py} -- Projected sensitivity plotting example script
\item \texttt{examples\_propagation.py} -- Waveform propagation example script
\item \texttt{*.ipynb} -- Jupyter notebooks for interactive analysis
\end{itemize}

\paragraph{\underline{inputs}: Input specification and configuration classes} This module contains the spectrum source hierarchy (\texttt{spectrum\_sources.py}), providing implementations for CSV-based spectra, analytical functions, and time-varying signals. The configuration classes (\texttt{configs.py}) encapsulate physics parameters (mass, couplings, burst duration, etc.) and propagation settings (matter density profile, interpolation resolution). The source-experiment-spectrum system (\texttt{source.py}, \texttt{experiment.py}, \texttt{spectrum.py}) represents the input configuration for projected sensitivity calculations, with \texttt{Source} containing astrophysical parameters, \texttt{Experiment} containing experimental parameters, and \texttt{SignalModel} performing the projected sensitivity computations.

\paragraph{\underline{waveform}: Numerical propagation workflow} The \texttt{collection.py} module provides the \texttt{WaveformCollection} class, which orchestrates the complete propagation process by combining a spectrum source with physics and propagation configurations. The \texttt{propagation.py} module implements the core numerical integration, computing spectrograms by propagating each frequency mode through the density profile. This module also provides plotting functions for visualizing spectrograms, as well as bridge functions to export results to the projected sensitivity plotting workflow.

\paragraph{\underline{plotting}: Visualization and output handling} The \texttt{plots.py} module defines plotting functions for individual projected sensitivity plot elements (time delay curves, screening regions, experimental limits, etc.) as well as styling and axis configuration. The \texttt{limits.py} module loads experimental constraint data from external files. The \texttt{OutputHandler} class in \texttt{output\_handler.py} manages figure creation for projected sensitivity plots, handling both single-panel and multi-panel grid layouts. It coordinates with the plotting functions in \texttt{plots.py} to apply consistent styling and manages the placement of exclusion regions, labels, and legends. 

\paragraph{\underline{utils}: Shared utilities} The \texttt{data\_utils.py} module provides functions for loading and interpolating CSV data (used for both spectrum files and density profiles). The \texttt{expt\_params.py} module stores experimental benchmark values (DM sensitivities, energy density fractions, astrophysical constraint benchmarks) and the file paths to external constraint data. The \texttt{validation.py} module provides helper functions used by the \texttt{Source} and \texttt{Experiment} constructors. The \texttt{logging\_utils.py} module configures logging with dual output to console and file, essential for tracking propagation progress and debugging.

\paragraph{\underline{calculations}: Analytical physics functions} The \texttt{calculations/sensitivities.py} module implements analytical formulas for coupling thresholds, time delays, screening couplings, and other sensitivity and constraint calculations, as well as the \texttt{propagation\_time()} function used by the waveform propagation workflow. The \texttt{utils/constants.py} module centralizes physical constants and unit conversions, ensuring consistency across all calculations.

\subsection{Source Spectrum Implementation}

A key code design element is the spectrum source abstraction, which allows the propagation workflow to operate on different input types through a common interface. The abstract class \texttt{SpectrumSource} defines three essential methods: \texttt{get\_spectrum()}, which returns arrays of momenta and amplitudes, \texttt{get\_num\_time\_steps()}, which returns the number of time steps (1 for a static spectrum), and \texttt{get\_time\_window()}, which returns the time window for time-dependent spectra. Concrete implementations are provided for different spectrum generation strategies. 

The \texttt{CSVSpectrum} class loads spectral data from CSV files produced by external simulations or theoretical calculations. It handles various file formats through configuration column indices and separators, and optionally interpolates data to a specified number of points for computational efficiency. It also supports both scalar and callable scaling functions for momentum and amplitude, enabling unit conversions and normalization without modifying the source files. 

The \texttt{AnalyticSpectrum} class generates spectra from analytical formulas, currently supporting Gaussian profiles, where users can specify the peak location, width, and amplitude.  The implementation ensures proper normalization, considering Gaussian sampled within few widths of peak, and samples the spectrum at energies relevant for propagation.

The \texttt{TimeVaryingSpectrum} class wraps another spectrum source with a time-dependent amplitude modulation function. The base spectrum provides the frequency structure, while the modulation function determines how the overall amplitude evolves with time. 

This architecture makes it straightforward to add new spectrum types by implementing the \texttt{SpectrumSource} interface without modifying any propagation code. That is, new CSV spectrum files or analytical functions can replace the built-in examples here.

\subsection{Configuration System}
Physics parameters and propagation settings are encapsulated in dedicated configuration classes for clarity. The \texttt{PhysicsConfig} class bundles the ULB field mass, coupling strength, energy density fraction, and burst duration. The \texttt{PropagationConfig} class specifies the path to the density profile and interpolation resolution.

This separation allows users to reuse the same spectrum with different physics configurations or to test different density profiles with the same source spectrum. Configuration objects are passed to the \texttt{WaveformCollection} class, which coordinates all aspects of the propagation calculation.

\subsection{Projected Sensitivity Calculation Workflow}
The projected sensitivity plotting workflow uses a different set of classes optimized for rapid analytical calculations. The \texttt{Source} class contains astrophysical source parameters: total emitted energy, ULB mass, burst duration, distance, and coupling configuration (type and order). The \texttt{Experiment} class encodes detector capabilities: integration times for both transient and background DM searches, fractional frequency sensitivity, and measurable time delay thresholds (a dictionary mapping labels to threshold values in seconds). 

These two objects are combined in the \texttt{SignalModel} class, which performs all projected sensitivity calculations by invoking analytical formulas from the \texttt{calculations/sensitivities.py} module. The \texttt{SignalModel} computes arrays of coupling strengths as functions of energy: the detection threshold coupling (below which signals are too weak to detect), time delay couplings (required to produce measurable delays), critical screening couplings (above which environmental screening suppresses the signal), and the energy-time uncertainty threshold. It also retrieves astrophysical constraints for the relevant couplings type.

The \texttt{OutputHandler} class takes a \texttt{Plot} configuration object and one or more \texttt{Source}-\texttt{SignalModel} pairs, creating matplotlib figures with appropriate axes, scales, and styling. For single sources, it produces a $1 \times 1$ plot; for 2D arrays of sources, it creates multi-panel grids showing how projected sensitivities vary with source parameters (typically mass and burst duration). The handler invokes plotting functions from \texttt{plots.py} to draw individual sensitivity and constraint elements (curves, shaded regions, labels), ensuring visual consistency across all panels. 

\subsection{Bridge Functions}
The two workflows (source propagation and projected sensitivity plotting) are connected through a set of bridge functions in \texttt{waveform/propagation.py} that convert between representations.

After running a detailed propagation simulation, users can call \texttt{export\_source\_parameters()} to record the source parameters and summary quantities derived from the signal at the detector (energy, mass, distance, burst duration) to a text file. 
This file serves as a record of the simulation and as input for projected sensitivity plotting. The notebook \texttt{examples\_complete\_workflow.ipynb} demonstrates this full pipeline end to end, from an input spectrum through propagation to a projected sensitivity plot.
  
  The approach uses the average detector energy density and the arrival  time interval to construct an effective energy normalization for the simplified sensitivity calculation. The normalization assumes relativistic propagation when converting a duration into a shell thickness and should not, in general, be identified with the physical emitted energy. The latter must be obtained from the source spectrum or, for the transmitted signal, from the velocity-weighted energy flux at the detector.

  \begin{figure*}[t]
    \centering \includegraphics[width = 0.9\linewidth]{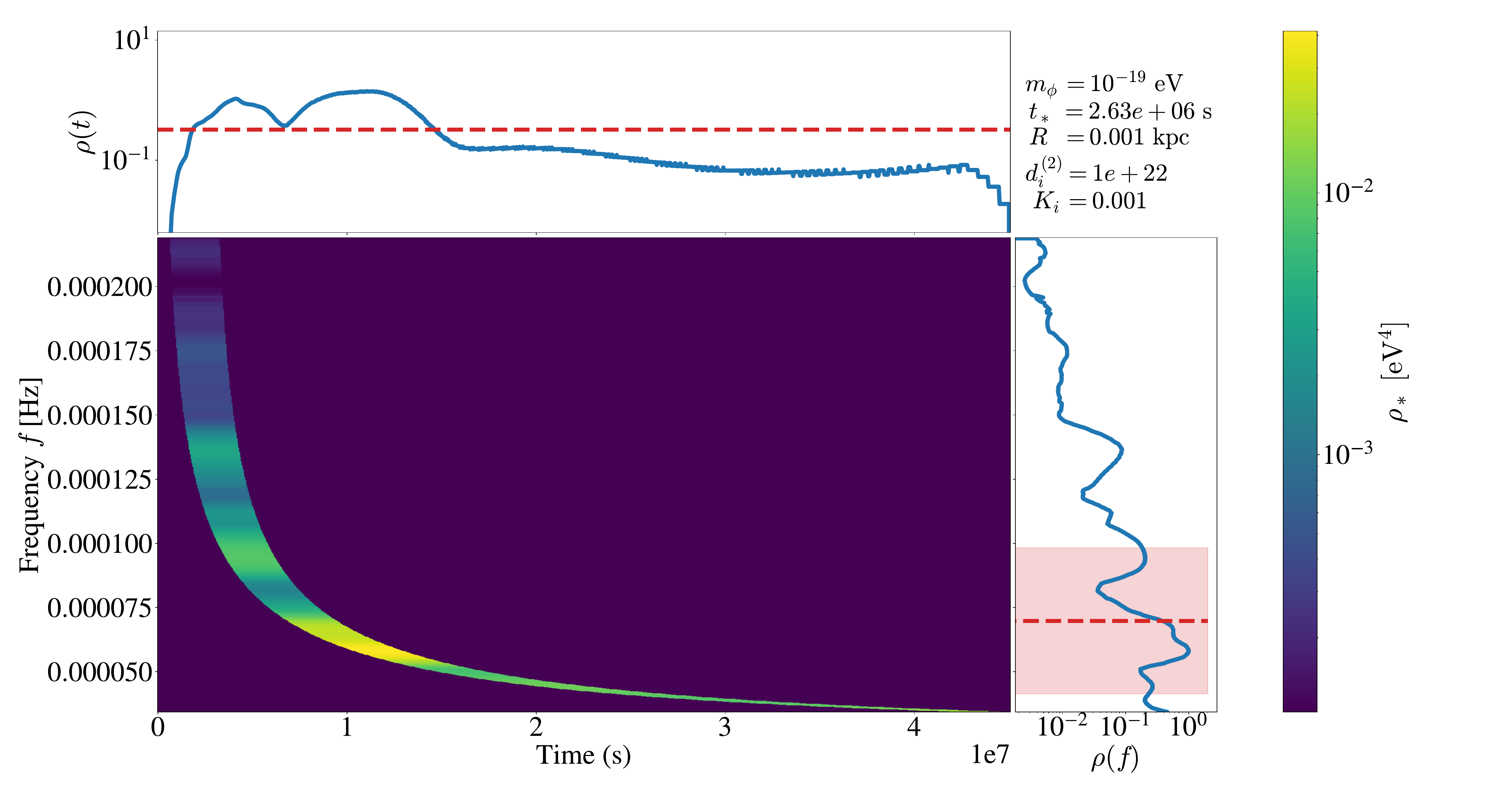}
    \caption{Spectrogram of a time-independent bosenova emission spectrum at Earth. This displays the frequency versus time dependence of the signal as it reaches the detector. The color bar represents the energy density of the field in each pixel in the time-frequency plane. The input parameters used here are displayed in the top right. The top panel represents the frequency-integrated energy density as a function of time, with the time-averaged energy density given by the red dashed line. The right panel shows the time-integrated frequency distribution, which is used to calculate the average frequency (red dashed) and standard deviation of frequencies (light red box).}
    \label{fig:bosenova_spectrogram}
\end{figure*}

\subsection{Units and Physics Constants}

All calculations use natural units with eV as the base energy unit. The \texttt{constants.py} module provides conversion factors between physical units and natural units: \texttt{SEC\_TO\_INEV} converts seconds to eV$^{-1}$, \texttt{PC\_TO\_INEV} converts parsecs to eV$^{-1}$, \texttt{SOLAR\_TO\_EV} converts solar masses to eV, and \texttt{GCM3\_TO\_EV4} converts densities from g~cm$^{-3}$ to eV$^4$. These conversions are applied consistently throughout the codebase, ensuring dimensional correctness, while allowing users to specify inputs in convenient physical units. 

Energy density fractions $K$ for different coupling types are stored in the \texttt{ENERGY\_DENSITY\_FRACTIONS} dictionary in \texttt{utils/expt\_params.py} \cite{Stadnik:2021qyf}, with distinct values for photon, electron, and gluon couplings reflecting the different energy content in these sectors. DM sensitivity benchmarks for each coupling type are provided in the same file's \texttt{DM\_SENSITIVITIES} dictionary, representing realistic detection thresholds for current and future experiments \cite{AxionLimits}.

\subsection{Data Files}
\label{sec:datafiles}
The repository includes necessary data files under \texttt{data/}:
\begin{itemize}[noitemsep]
    \item \texttt{Galactic\_Density\_Profile.csv}: Galactic disk density profile $\rho(r)$ derived from observational data~\cite{Nakanishi_2015}.
    \item \texttt{Linear Scalar Photon/}, \texttt{Linear Scalar Electron/}, \texttt{Linear Scalar Gluon/}: Constraint data from laboratory experiments (MICROSCOPE \cite{Berge:2017ovy}, fifth force~\cite{Adelberger:2003zx,Fischbach:1996eq, KONOPLIV2011401}, E\"ot-Wash\cite{Hees:2018fpg}) for the linear scalar photon, electron, and gluon couplings, respectively. MICROSCOPE bounds are provided for all three coupling types. Fifth force and E\"ot-Wash bounds are provided for the photon and electron couplings only. The E\"ot-Wash files are included for completeness, but not currently read by the plotting code.
    \item \texttt{BosonStarSpectrumRelOnly.txt}: Example bosenova emission spectrum file (see Sec.~\ref{sec:bosenova}).
    \item \texttt{Source@R=1parsec.csv}: Density profile for a source 1~pc from Earth, in the same format as \texttt{Galactic\_Density\_Profile.csv}.
\end{itemize}

The density profile is formatted in the following way: column 1 is the position in kiloparsecs (kpc), column 2 is the number density of particles in the medium in cm$^{-3}$, and column 3 is the energy density of those particles in g~cm$^{-3}$. In the case of the provided file \texttt{Galactic\_Density\_Profile.csv}, the particles are hydrogen species, which allows us to convert from number density to mass density. The first and last entry in the first column should represent the position of the source and the Earth respectively.

Users can substitute a custom density profile by providing a file in the same format and passing its path directly to \texttt{PropagationConfig}. Constraint data can similarly be substituted by providing files in the same format, and updating the corresponding file paths in \texttt{utils/expt\_params.py}. The default list of exclusion limits used here were taken from \cite{AxionLimits}.

\begin{table*}[t]
    \centering
    \begin{tabular}{c|c|c|c}
        Spectrum Type & Spectrum Integration at Source (eV) & Spectrogram Integration at Earth (eV) & Fractional Difference\\
        \hline
        Time-indep. Gaussian &  $1.08766\times 10^{66}$  & $1.08770 \times 10^{66}$ & $3.5\times10^{-5}$\\
        Time-dep. Gaussian &  $6.1912\times 10^{65}$ & $6.1907\times10^{65}$ & $7.2\times 10^{-5}$\\
        Bosenova & $6.1093\times 10^{69}$ & $6.1036\times 10^{69}$ & $9.3 \times 10^{-4}$\\
    \end{tabular}
    \caption{Comparisons of energy calculated at source and at Earth. The table gives values for time-independent and time-dependent Gaussian spectra, as well as for the bosenova example spectrum discussed in Sec.~\ref{sec:bosenova}. All entries use $N_t = 1000$, $N_p = 1000$, $d_i^{(2)} = 10^{22}$, and $K = 10^{-3}$. The methods agree to differences of less than a percent for the resolutions used here.}
    \label{tab:energyconservation}
\end{table*}

\subsection{Spectrum CSV Input File}

Users can provide a specific spectrum input file, giving the code the distribution of energy per momentum bin as a function of the emission momentum. The CSV file contains two columns, with the first column being the unitless ratio $p/m$ where $m$ and $p$ are the momentum of the emitted field, and the second the unitless differential energy $(dE/dp)/A$ scaled by a unitless normalization constant $A$. The normalization of $(dE/dp)/A$ can be chosen by the user. To be explicit, the relation between the second column and the energy spectrum is 

\begin{align}
    \frac{dE}{dp} = A \cdot ({\rm column~2})~.
\end{align}

In the case that $A = 1$, the second column is exactly the differential energy spectrum $dE/dp$. The explicit value of $A$ can be chosen in the code, external to the CSV file, using the \texttt{amplitude\_normalization} line in the code (see examples below in Sec.~\ref{sec:bosenova}).

It is also possible to interpolate between points in the CSV file to more finely bin the momenta than the CSV file provides. When loading the CSV file, the user can call linear or logarithmic spacing for the points for the interpolation, depending on which is more useful given the input.

\section{Bosenova Spectrum Example}
\label{sec:bosenova}

We demonstrate the utility of our code on the example of bosenova emission.
A boson star is a gravitationally bound configuration of ULB fields. In the presence of ULB self-interactions, collapse and explosion of boson stars can result in significant emission of relativistic ULBs that we consider to be symmetrically emitted in a spherical shell. This process, known as a bosenova, has been simulated in Ref.~\cite{Levkov:2016rkk}, providing a concrete emission spectrum. 

The bosenova example applies the propagation methodology of Sec.~\ref{sec:method} to the emission spectrum calculated in Ref.~\cite{Levkov:2016rkk}, rather than an idealized Gaussian signal. The momentum  spectrum is loaded from a CSV file produced by digitizing Fig.~3(a) in Ref.~\cite{Levkov:2016rkk} using \texttt{WebPlotDigitizer} \cite{githubGitHubAutomerisioWebPlotDigitizer}. The spectrum has a normalization factor that corresponds to a self-interaction coupling $\lambda$, such that $A = 1/\lambda$. In this way, a single spectrum CSV file can be used for any choice of $\lambda$. For a given mass $m$, there is a range of values that $\lambda$ can take that is consistent with observation. This can be seen in Fig. 1 of Ref.~\cite{Arakawa:2023gyq}. We choose a value $\lambda = 10^{-85}$ as a benchmark.

The parameters a user must supply are:
\begin{itemize}
    \item \texttt{mass} -- ULB field mass $m$ in eV.
    \item \texttt{burst\_duration\_factor} -- sets the burst duration via $t_* = \texttt{burst\_duration\_factor}/(m\cdot\texttt{SEC\_TO\_INEV})$.
    \item \texttt{amplitude\_normalization} -- overall rescaling of the CSV amplitude column corresponding to $1/\lambda$ (e.g.\ $1/10^{-85}$ for the benchmark $\lambda = 10^{-85}$ used here).
    \item \texttt{coupling}, \texttt{K} -- dilatonic coupling strength and energy density fraction.
    \item Path to the spectrum CSV file (\texttt{data/BosonStarSpectrumRelOnly.txt}) and the column indices \texttt{i\_p}, \texttt{i\_A} for momentum and amplitude.
    \item \texttt{scaled\_momentum} and \texttt{scaled\_amplitude} -- rescaling factors that convert the dimensionless CSV data to physical units, here multiplying momenta by \texttt{mass} and amplitudes by \texttt{amplitude\_normalization}.
    \item \texttt{density\_num\_points} -- interpolation resolution for the density profile 
    \item \texttt{num\_points} -- number of new grid points for interpolation of the CSV spectrum file.
\end{itemize}

The spectrogram of a bosenova occurring at distance of 1 parsec  away from Earth is shown in Fig.~\ref{fig:bosenova_spectrogram}. Since the bosenova emission is semi-relativistic with an order of magnitude momentum spread, the wave spreading effect extends the signal to have a very long duration of the order of years. The slope of the spectrogram shows that the signal is broader in frequency early on, becoming nearly monochromatic with time. The duration of the signal here is consistent with the analytical approximations for the duration calculated in Ref.~\cite{Eby:2021ece}. Additionally, the spectrogram shows the non-trivial, multi-peak shape of the bosenova spectrum, with the maximum amplitude signal arriving at $t \simeq 10^{7}$ s.

\section{Code Validation}

\subsection{Energy Conservation}

We have verified that over the propagation the total emitted ULB energy   remains conserved. That is, integrating the density of the field over the duration of the signal at Earth and the spherical emission of radius $r$ matches the integration of the input spectrum $dE/dp$ over momentum 
\begin{align}
    \int \frac{dE}{dp} dp \bigg|_{\rm Source} \simeq 4\pi r^2 \int \sum_i \rho_i(t)\, v_{i}\,dt \bigg|_{\rm Earth}
\end{align}
to sufficient accuracy.  For the parameters chosen in the examples, we find that energy is generally conserved to less than a percent, for the resolutions used here, as shown in Tab.~\ref{tab:energyconservation}.  For screened signals, the energy comparison includes only modes transmitted to the detector.

\subsection{Convergence}

The accuracy and precision of the results ultimately depend on the degree to which the discrete binning approximates the spectra and timing. In Tab.~\ref{tab:convergence} we find that for all cases considered the fractional difference between the source spectrum and detector flux energy integrals decreases as the momentum and time resolutions are increased for the bosenova benchmark.

\subsection{Propagation time}

We have verified that the code converges to the analytical calculation of the propagation time in the constant density limits. As discussed above, the time of arrival is given by Eq.~\eqref{eq:t_arrival}. See Tab.~\ref{tab:arrival_times}. The arrival times using the numerical integration versus the analytical values match to high degree of accuracy.

\section{Acknowledgments}
\label{sec:acknowledgments}

We thank Kira Scharf and Arturo de Giorgi for useful discussions. This work was supported by the NSF Quantum Leap Challenge Institute (Award OMA-2016244), the NSF award 2326810, and the European Research Council (ERC) under the European Union's Horizon 2020 research and innovation program (Grant No. 856415). M. H. Z. acknowledges support of the NSF Grant No. 2608275. This research was supported in part through the use of University of Delaware HPC Caviness and DARWIN computing systems. 
J. A. acknowledges hospitality at WPI-QUP, KEK during a visit hosted by V. Takhistov. This work was supported by the
World Premier International Research Center Initiative
(WPI), MEXT, Japan. V.T. acknowledges support in part by the MEXT SPReAD Grant No. 26304205.

\begin{table}[t]
    \centering
    \begin{tabular}{c|c|c|c}
         & $N_t = 100$ & $N_t = 1000$ & $N_t = 10000$\\
        \hline
        $N_p = 100$  & $1.2 \times 10^{-2}$ & $5.3 \times 10^{-3}$ & $3.6 \times 10^{-3}$ \\
        $N_p = 1000$ & $3.0 \times 10^{-3}$  & $9.4 \times 10^{-4}$ & $3.3\times10^{-4}$ \\
        $N_p = 10000$ & $7.1 \times 10^{-4}$ & $7.2 \times 10^{-5}$  & $2.0 \times 10^{-5}$ \\
    \end{tabular}
    \caption{Demonstration of convergence of fractional difference between the bosenova energy integrated using the source spectrum and the spectrogram. We choose different values of the number of grid points $N_t$ and $N_p$. All entries use $d_i^{(2)} = 10^{22}$, and $K = 10^{-3}$.}
    \label{tab:convergence}
\end{table} 

\begin{table}[t]
    \centering
    \begin{tabular}{c|c|c|c}
       & GW & No screening & Constant Density\\
       \hline
       $t_{\rm arrival, analytical}$ [yr]  & 3.26726 & 3.28736 & 4.02086 \\
       $t_{\rm arrival, numerical}$ [yr]&  --  & 3.28736  &  4.02086 \\
    \end{tabular}
    \caption{Comparison of the explicit analytical time of arrival for the fastest momentum mode ($\omega = 9.05 \times 10^{-19}$ eV for $m = 10^{-19}$ eV) for a bosenova source at $R \simeq 1$~pc compared to the arrival time generated by the code, with respect to gravitational wave emission (GW). They match exactly up to rounding, demonstrating that the numerical integration over a constant density profile is working as expected. For the constant density screening results, we use a coupling $d_i^{(2)} = 5\times10^{25}$, and $K = 10^{-3}$.}
    \label{tab:arrival_times}
\end{table}

\appendix

\section{Particle Physics Description}
\label{app:particlephysics}

We summarize the field interactions relevant to the propagation and
detection calculations. Both scalar fields and pseudoscalar ALPs are denoted by $\phi$, with the
field type specified where needed.
The free field Lagrangian is
\begin{align}
    \mathcal{L}_{\phi}
    = \frac{1}{2}\partial_{\mu}\phi\,\partial^{\mu}\phi
    - \frac{1}{2}m^2\phi^2~.
\end{align}
This gives the Klein-Gordon equation, whose plane wave modes
provide the basis for the propagation calculation.

The interactions considered here have the schematic forms
\begin{align}
    \mathcal{L}_{\rm int}^{\rm linear}
    &\sim \phi\,\mathcal{O}_{\rm SM}~,\\
    \mathcal{L}_{\rm int}^{\rm quad.}
    &\sim \phi^2\,\mathcal{O}_{\rm SM}~,\\
    \mathcal{L}_{\rm int}^{\rm deriv.}
    &\sim \partial_{\mu}\phi\,\mathcal{O}^{\,\mu}_{\rm SM}~,
\end{align}
where the operators are constructed from Standard Model  
fields. Examples include $m_e\bar{e}e$,
$F_{\mu\nu}F^{\mu\nu}/4$, and the fermion axial current
$\bar{f}\gamma^\mu\gamma^5 f$, whose spatial components
describe spin density in the nonrelativistic limit.
For parity-conserving interactions, the operators accompanying
$\phi$ and $\partial_\mu\phi$ differ between scalars and
pseudoscalars. Quadratic interactions can have the same
operator structure because $\phi^2$ is parity even in both
cases.

\subsection{Scalars}

We consider linear and quadratic scalar couplings to electrons,
photons, and gluons 
\begin{align}
\label{eq:scalar_Lint}
    \mathcal{L}^{(s)}_{\text{int}}
    &=
    \frac{\sqrt{4\pi}\,\phi}{\Mpl}
    \bigg(
        -d_{m_e}^{(1)}m_e\overline{e}e
        +\frac{d_e^{(1)}}{4}F_{\mu\nu}F^{\mu\nu}
        +\frac{d_g^{(1)}\beta(g_s)}{2g_s}
        G_{\mu\nu}^aG^{a\mu\nu}
    \bigg)
    \nonumber\\
    &\quad+
    \frac{4\pi\,\phi^2}{\Mpl^2}
    \bigg(
        -d_{m_e}^{(2)}m_e\overline{e}e
        +\frac{d_e^{(2)}}{4}F_{\mu\nu}F^{\mu\nu}
        +\frac{d_g^{(2)}\beta(g_s)}{2g_s}
        G_{\mu\nu}^aG^{a\mu\nu}
    \bigg)~.
\end{align}
Here $\Mpl=1.2\times10^{28}~\mathrm{eV}$ is the Planck mass,
$m_e$ and $e$ are the electron mass and field,
$F_{\mu\nu}$ and $G_{\mu\nu}^a$ are the electromagnetic and
gluon field strengths, and $a$ is a color index.
The strong coupling is $g_s$, and $\beta(g_s)$ is its QCD
beta function. The dimensionless coefficients $d_i^{(n)}$
specify the interaction strengths, with $n=1,2$ denoting
linear and quadratic couplings.

These interactions produce field-dependent shifts in the
electron mass, fine-structure constant, and QCD scale.
For each coupling order, the corresponding fractional
variations are written as
\begin{align}
\label{eq:deltafundconst}
    \frac{\delta m_e}{m_e}(t)
    &=
    d_{m_e}^{(n)}
    \left(\frac{\sqrt{4\pi}\,\phi(t)}{\Mpl}\right)^n~,
    \nonumber\\
    \frac{\delta\alpha}{\alpha}(t)
    &=
    d_e^{(n)}
    \left(\frac{\sqrt{4\pi}\,\phi(t)}{\Mpl}\right)^n~,
    \nonumber\\
    \frac{\delta(m_q/\Lambda_{\rm QCD})}
         {m_q/\Lambda_{\rm QCD}}(t)
    &=
    d_g^{(n)}
    \left(\frac{\sqrt{4\pi}\,\phi(t)}{\Mpl}\right)^n~.
\end{align}
The last relation expresses the gluon-induced response using
the sign convention of Eq.~\eqref{eq:scalar_Lint}, with
the quark masses held fixed. Observable mass ratios, such as
$m_e/m_p$, depend on the responses of both masses.

\subsection{Axion-like Particles}

For pseudoscalar ALPs, we consider
\begin{align}
\label{eq:ALP_Lint}
    \mathcal{L}_{\rm int}^{(\rm ps)}
    &=
    \phi\big(
        g_{\phi\gamma\gamma}F_{\mu\nu}\tilde{F}^{\mu\nu}
        +g_{\phi gg}G_{\mu\nu}^a\tilde{G}^{a\mu\nu}
    \big) +
    g_{\phi f}(\partial_\mu\phi)
    \overline{f}\gamma^\mu\gamma^5 f~,
\end{align}
where $f$ denotes a Standard Model fermion, including electrons and
quarks. The dual field strengths are
$\tilde{F}^{\mu\nu}
=\frac{1}{2}\epsilon^{\mu\nu\rho\sigma}F_{\rho\sigma}$
and
$\tilde{G}^{a\mu\nu}
=\frac{1}{2}\epsilon^{\mu\nu\rho\sigma}G_{\rho\sigma}^a$.

At the low energies relevant to the detectors considered
here, quark and gluon interactions are expressed in terms
of nucleon and meson degrees of freedom. Matching to
hadronic interactions can be performed using chiral
perturbation theory, see for example Ref.~\cite{Scherer:2002tk}.
The relevant nucleon interactions can be parameterized as
\begin{align}
\label{eq:ALP_nucl}
    \mathcal{L}_{\rm nuc}^{(\rm ps)}
    &=
    g_{\phi\mathcal{N}\mathcal{N}}
    (\partial_\mu\phi)
    \bar{\mathcal{N}}\gamma^\mu\gamma^5\mathcal{N} -
    \frac{i}{2}g_{\phi\mathcal{N}\gamma}\phi
    \bar{\mathcal{N}}\sigma_{\mu\nu}\gamma_5\mathcal{N}
    F^{\mu\nu}~,
\end{align}
where $\mathcal{N}$ denotes a proton or neutron.
The first term gives a coupling to nucleon spin, while the
second describes a $\phi$-dependent nucleon electric
dipole moment.

\subsection{In-medium Effects}

We describe the medium through expectation values of the
Standard Model operators entering the quadratic interactions. For the
scalar couplings above, define
\begin{align}
\label{eq:expenv}
    \rho_i
    &= \langle\rho|\mathcal{O}_{\rm SM}^i|\rho\rangle
    \nonumber\\
    &=
    \begin{cases}
        \langle\rho|m_e\bar{e}e|\rho\rangle
        & \text{electrons},\\
        \langle\rho|-F_{\mu\nu}F^{\mu\nu}/4|\rho\rangle
        & \text{photons},\\
        \langle\rho|-[\beta(g_s)/(2g_s)]
        G_{\mu\nu}^aG^{a\mu\nu}|\rho\rangle
        & \text{gluons},
    \end{cases}
\end{align}
where $|\rho\rangle$ denotes the state of the medium.
These expectation values are parameterized by
$K_i=\rho_i/\rho_{\rm medium}$.
For hydrogen gas, for example, the electron rest mass
contribution is approximately
$\rho_e\simeq5\times10^{-4}\rho_H$.

Replacing the Standard Model operators by their medium expectation
values gives a contribution to the $\phi$ mass term.
For electrons,
\begin{align}
    -\frac{4\pi}{\Mpl^2}
    d_{m_e}^{(2)}m_e\bar{e}e\,\phi^2
    \;\longrightarrow\;
    -\frac{4\pi}{\Mpl^2}
    d_{m_e}^{(2)}\rho_e\,\phi^2~.
\end{align}
Defining
$\beta=\sum_i8\pi d_i^{(2)}\rho_i/\Mpl^2$
and including a quartic self-interaction gives
\begin{align}
\label{eq:L_with_beta}
    \mathcal{L}
    \supset
    \frac{1}{2}\partial_\mu\phi\,\partial^\mu\phi
    -\frac{1}{2}(m_\phi^2+\beta)\phi^2
    -\frac{\lambda}{4!}\phi^4~.
\end{align}
The self-interaction coefficient $\lambda$ may have either
sign; the attractive case is discussed in the bosenova
appendix. The numerical propagation calculation uses
the effective mass $m_{\rm eff}^2=m_\phi^2+\beta$ and
does not evolve the quartic self-interaction.
Modes with $\omega^2>m_{\rm eff}^2$ propagate with the
corresponding group velocity, while modes encountering
$\omega^2\leq m_{\rm eff}^2$ are treated as screened.

\section{Emission Sources}
\label{app:sources}

Predicting a transient ULB signal at a detector requires specifying its emission at the source, propagation to Earth, and detector response. In the context of multimessenger searches, the source is considered to also produce an observable signal in Standard Model and known messengers, such as electromagnetic radiation, neutrinos or gravitational waves. Here we summarize the source parameters, propagation through the intervening medium  and simplified detection estimates used in this work. The source-to-detector framework underlying this discussion was introduced in Ref.~\cite{Eby:2021ece}, with the in-medium multimessenger treatment further developed in Ref.~\cite{Arakawa:2025hcn}.

The code supports two descriptions of relativistic ULB sources. A parameterized description specifies the total emitted energy, characteristic frequency, spectral width, burst duration and source distance for analytical estimates of the signal at Earth. This is useful for parameter scans or when a detailed emission spectrum is unavailable. Alternatively, a supplied emission spectrum, such as from simulations, can be propagated numerically to obtain the time-frequency energy density distribution at Earth. We describe both approaches below.

\subsection{Source Characterization}

When a detailed emission spectrum is unavailable, we can model  
the source in simplified way with general phenomenological parameters. This includes the total energy
emitted in $\phi$, $E_{\rm tot}$, the mean particle energy 
$\omega_\phi$, the energy spread $\delta\omega$, the intrinsic
burst duration $t_*$, and the source-to-Earth distance  $r$.
These parameters allow an estimate of the field amplitude at
Earth using Eqs.~\eqref{eq:phi0} and \eqref{eq:rho_approx} of
Sec.~\ref{sec:theory}, without specifying the full emission
spectrum $dE/dp$. This parameterized description underlies the
\texttt{Source} object used in the projected sensitivity
plotting workflow (Sec.~\ref{sec:sensitivity}).

Alternatively, the source can be specified by an emission
differential energy spectrum $dE/dp$, together with its burst duration
and distance. The contribution to the energy density at Earth
from a momentum bin centered at
$p_i=\sqrt{\omega_i^2-m^2}$ with width $\Delta p$ is given by
Eq.~\eqref{eq:rho_pt}. Its normalization uses the per-mode
shell thickness
$x_{*,i}=v_g(p_i)t_*=\sqrt{1-m^2/\omega_i^2}\,t_*$,
evaluated with the vacuum group velocity at the observation
point. The overall spreading of the signal arises from the
different propagation times of the momentum modes, as described
in Sec.~\ref{sec:theory} and implemented numerically in
Sec.~\ref{sec:method}.

\subsection{Bosenovae}
 
Boson stars are gravitationally bound configurations of ULBs,
which can arise in context of ultralight DM. In the dilute regime, they are supported by a balance
between gradient energy and self-gravity. Scalar
self-interactions can modify this balance and lead to collapse.
The source dynamics considered here involve gravity and
interactions within the $\phi$ sector. Standard Model couplings enter
the subsequent propagation and detection calculations.

We consider the Lagrangian
\begin{align}
    \mathcal{L}_{\phi}
    = \frac{1}{2}\partial_{\mu}\phi\,\partial^{\mu}\phi
    - \frac{1}{2}m_{\phi}^2\phi^2
    + \frac{\lambda}{4!}\phi^4 + \dots~,
\end{align}
where $\lambda>0$ denotes the magnitude of the attractive
quartic self-interaction. This sign convention differs from
Eq.~\eqref{eq:L_with_beta}. The ellipsis denotes higher-order
self-interactions, which are not specified here. Constraints
on $\lambda$ as a function of $m_\phi$ are summarized in
Fig.~1 of Ref.~\cite{Arakawa:2023gyq}.

In the nonrelativistic limit, the field is described by a
condensate wavefunction $\psi$ obeying the
Gross-Pitaevskii equation coupled to Newtonian gravity 
\begin{align}
    \left(
        -\frac{\nabla^2}{2m_\phi}
        + V_g(|\psi|^2)
        - \frac{\lambda}{8m_\phi^2}|\psi|^2
    \right)\psi
    = i\dot{\psi}~,
\end{align}
with
\begin{align}
    \phi
    = \frac{1}{\sqrt{2m_\phi}}
    \left(
        \psi e^{-im_\phi t}
        + \psi^*e^{im_\phi t}
    \right)~.
\end{align}
Here $V_g=m_\phi\Phi_g$ is the gravitational potential energy
per particle, with the Newtonian potential $\Phi_g$ determined
by the condensate density through Poisson's equation.

Boson stars can gain mass, such as by accretion or
merging with other boson stars. As the density increases,
the attractive self-interaction becomes more significant,
although the coupling $\lambda$ itself remains fixed.
The dilute equilibrium becomes unstable near the critical mass
$M_{\rm crit}\simeq10M_{\rm pl}/\sqrt{\lambda}$
\cite{Arakawa:2023gyq,Arakawa:2024lqr}, leading to collapse.
In the bosenova regime considered here, relativistic
field dynamics interrupt the collapse and produce outgoing
ULBs rather than a black hole. These dynamics include
number-changing processes such as $3\phi\rightarrow\phi$
in the spatially varying condensate, which transfer energy
to escaping relativistic modes. This results in a bosenovae emission. In the simulations considered in
Ref.~\cite{Levkov:2016rkk}, the collapse and emission sequence
ejects approximately $30\%$ of the initial mass, leaving
a less massive bound remnant.

We use the rescaled emission spectrum from
Ref.~\cite{Levkov:2016rkk}, stored as a function of $p/m_\phi$
in the supplied CSV file. Its normalization is set by
$\lambda$ 
\begin{align}
    \frac{dE}{dp}
    = \frac{\text{CSV column 2}}{\lambda}~.
\end{align}

\section{Analytical Propagation Estimates}
\label{app:prop_analytical}

 Using the arrival time defined in Eq.~\eqref{eq:t_arrival}, we can provide an analytical estimate of propagation under simplifying assumptions, following the procedure outlined in Ref.~\cite{Arakawa:2025hcn}. In general, the arrival time for quadratically-coupled fields $\phi$ depends on the distribution of SM matter throughout the propagation. For an extragalactic source, the propagating $\phi$ wave may encounter multiple galaxies on its path towards Earth, requiring information about both the galaxy distribution and matter distribution within galaxies. Due to the complicated nature of the distribution of matter within each galaxy, we focus only on Milky Way-type galaxies and assume for simplicity that each has similar densities of gas and dust in their respective interstellar media (ISM), and that their distribution is homogeneous within each galaxy. Additionally, $\phi$ traverses  intergalactic medium (IGM) in between galaxies. We consider the typical ISM and the IGM particle densities~(e.g.~\cite{Nakanishi_2015}) of $\sim 1$~cm$^{-3}$ and $\sim 1$~m$^{-3}$, respectively, with $\rho_{\rm ISM} \simeq 1.67 \times 10^{-24}~{\rm g~cm}^{-3}$  and $\rho_{\rm IGM} \simeq 1.67 \times 10^{-30}~{\rm g~cm}^{-3}$ assuming predominantly hydrogen gas for both.

For the analytical estimates, we parameterize the total path length through galactic matter considering effective galaxy count parameter $N_g$   determined by the population distribution (a volumetric number density) of galaxies $n(M_{\rm g})$, which depends on galaxy mass $M_{\rm g}$. That is, $N_g(M_{\rm g}) \simeq (n(M_{\rm g}))^{1/3}r$, where $r$ is the distance between the source and Earth. The number density of Milky Way-type galaxies is given by the Schechter function $n_{\rm MW} \sim \mathcal{O}(10^6)~{\rm Gpc}^{-3}$ \cite{Schechter}. We assume this as an effective phenomenological propagation benchmark.

The interaction length through each galaxy changes with the orientations of each galaxy, which, without specific a priori knowledge about the source location, is randomized.  For the projected sensitivity plots, we vary the assumed interaction length per galaxy between perpendicular and parallel disk-crossing benchmarks. The resulting band illustrates dependence on the adopted propagation geometry. The first is where $\phi$ enters every galaxy completely perpendicularly to the disk, and therefore propagates the shortest distance through the galaxy. Second is where the ULB enter every galaxy parallel to the galactic plane, representing the largest possible propagation distance through the galaxies. This propagation distance, or interaction length, through each galaxy is labeled $D_g$. Users can define their own $D_g$ rather than showing the range for these extrema defined here.

Altogether, the distance that the ULBs travel through galaxies is given by $N_g D_g$. Then, the integral for the arrival time can be split up piecewise, given that we approximated the densities in the ISM and IGM to be constant. Eq.~\eqref{eq:t_arrival} becomes \cite{Arakawa:2025hcn}, 

\begin{align} \label{eq:tphi}
    t_{\phi}
     =&~ \int_{0}^{N_g D_g}dx~  {\left(1- \frac{1}{\omega^2}\left(m_{\phi}^2 + \dfrac{8\pi\,d_i^{(2)} \rho_{{\rm ISM},i}}{\Mpl^2}\right)\right)}^{-1/2} \nonumber \\
    &+\int_{N_g D_g}^{r}dx~ ~ {\left(1- \frac{1}{\omega^2}\left(m_{\phi}^2 + \dfrac{8\pi\,d_i^{(2)} \rho_{{\rm IGM},i}}{\Mpl^2}\right)\right)}^{-1/2} ~.
\end{align}

The extragalactic time delay curves in the analytical sensitivity module are obtained using the ultrarelativistic expansion of Eq.~\eqref{eq:tphi}. The numerical propagation module instead evaluates Eq.~\eqref{eq:t_arrival} through the specified density profile without this expansion.  

\section{Field Coupling Detection Sensitivity}
\label{app:detection}

We estimate coupling sensitivities by rescaling the response
to a persistent nonrelativistic DM field to that of a transient
ULB signal, following procedure of
Refs.~\cite{Eby:2021ece,Arakawa:2025hcn}.
For estimates based on numerical propagation  the spectrogram
provides the average energy density, characteristic frequency,
and arrival window duration. The parameterized workflow
instead estimates the signal properties analytically.
The rescaling accounts for the finite signal duration and
the difference between the transient coherence time $\tau_*$
and the DM coherence time $\tau_{\rm DM}$.

For scalar ULBs with dilatonic couplings, the sensitivity
depends on the field amplitude $\phi_0$
(Eq.~\eqref{eq:phi0}) and the induced variations of fundamental
constants (Eq.~\eqref{eq:deltafundconst}). The estimated
thresholds for linear and quadratic couplings are
\cite{Eby:2021ece,Arakawa:2025hcn}
\begin{align}
\label{eq:d_star}
    d_{i,*}^{(n)} = \eta \left(\frac{M_{\rm pl}}{2\sqrt{\pi}\,\phi_0}\right)^{n} \frac{t_{\rm int,\, DM}^{1/4}
        {\rm min}\left(\tau_{\rm DM}^{1/4},t_{\rm int,\, DM}^{1/4}\right)}
        {{\rm min}\left(\tilde{t}_*^{1/4},t_{\rm int, *}^{1/4}\right)
        {\rm min}\left(\tau_*^{1/4},t_{\rm int, *}^{1/4}\right)}~,
\end{align}
where $\eta$ is the reference fractional sensitivity of the
detector, associated with the DM integration time
$t_{\rm int,\,DM}$.
The transient search uses an integration time $t_{\rm int,*}$,
and $\tilde{t}_*$ is the signal duration at the detector.
The index $n=1,2$ denotes linear and quadratic couplings,
respectively, as defined in
 \ref{app:particlephysics}.

For ALPs, we rescale a reference DM coupling sensitivity
$g_{\rm DM}$ according to
\cite{Eby:2021ece,Arakawa:2025hcn}
\begin{align}
    g_*  = g_{\rm DM}\sqrt{\frac{\rho_{\rm DM}}{\rho_*}}\bigg(\frac{v_{\rm DM}}{v_{*}}\bigg)^j  \frac{t_{\rm int,\, DM}^{1/4}
        {\rm min}\left(\tau_{\rm DM}^{1/4},t_{\rm int,\, DM}^{1/4}\right)}
        {{\rm min}\left(\tilde{t}_*^{1/4},t_{\rm int, *}^{1/4}\right)
        {\rm min}\left(\tau_*^{1/4},t_{\rm int, *}^{1/4}\right)}
\end{align}
where $\rho_{\rm DM}$ and $\rho_*$ are the reference DM
and transient energy densities, and $v_{\rm DM}$ and $v_*$
are their respective velocities. The exponent $j$ accounts
for the velocity dependence of the detector response.
The code assigns $j=1$ to the fermion spin couplings
$\sim(\partial_\mu\phi)\bar{\psi}\gamma^\mu\gamma^5\psi$
and $j=0$ to the other ALP couplings considered here.

The reference sensitivity and rescaling factors should be
evaluated at the same signal frequency. In natural units,
the relativistic field has angular frequency
$\omega_{\rm rel}=\sqrt{p^2+m_\phi^2}$, whereas a
nonrelativistic DM field has
$\omega_{\rm DM}\simeq m_{\rm DM}$.
Thus the comparison uses
$\omega_{\rm rel}=\omega_{\rm DM}$, corresponding to
$m_{\rm DM}\simeq\sqrt{p^2+m_\phi^2}$ rather than
$m_{\rm DM}=m_\phi$.

These expressions provide approximate sensitivity estimates.
An experiment specific analysis would combine the
time-frequency energy density distribution with the detector
response, noise properties, observation window and detection
statistic. Such analyses are outside the scope of this work.
The numerical propagation and spectrogram calculation do not
depend on the sensitivity rescaling prescription and can
provide inputs to separate detector analysis pipelines.
 
\bibliographystyle{elsarticle-num}
\bibliography{ref}

@article{Takahashi:2024fyq,
    author = "Takahashi, Takuya and Omiya, Hidetoshi and Tanaka, Takahiro",
    title = "{Self-interacting axion clouds around rotating black holes in binary systems}",
    eprint = "2408.08349",
    archivePrefix = "arXiv",
    primaryClass = "gr-qc",
    reportNumber = "RUP-24-12",
    doi = "10.1103/PhysRevD.110.104038",
    journal = "Phys. Rev. D",
    volume = "110",
    number = "10",
    pages = "104038",
    year = "2024"
}

@article{Raffelt:1999tx,
    author = "Raffelt, Georg G.",
    title = "{Particle physics from stars}",
    eprint = "hep-ph/9903472",
    archivePrefix = "arXiv",
    doi = "10.1146/annurev.nucl.49.1.163",
    journal = "Ann. Rev. Nucl. Part. Sci.",
    volume = "49",
    pages = "163--216",
    year = "1999"
}

@article{Baumann:2021fkf,
    author = "Baumann, Daniel and Bertone, Gianfranco and Stout, John and Tomaselli, Giovanni Maria",
    title = "{Ionization of gravitational atoms}",
    eprint = "2112.14777",
    archivePrefix = "arXiv",
    primaryClass = "gr-qc",
    doi = "10.1103/PhysRevD.105.115036",
    journal = "Phys. Rev. D",
    volume = "105",
    number = "11",
    pages = "115036",
    year = "2022"
}

@article{Yoshino:2012kn,
    author = "Yoshino, Hirotaka and Kodama, Hideo",
    title = "{Bosenova collapse of axion cloud around a rotating black hole}",
    eprint = "1203.5070",
    archivePrefix = "arXiv",
    primaryClass = "gr-qc",
    reportNumber = "KEK-TH-1530",
    doi = "10.1143/PTP.128.153",
    journal = "Prog. Theor. Phys.",
    volume = "128",
    pages = "153--190",
    year = "2012"
}

@article{Levkov:2016rkk,
    author = "Levkov, D. G. and Panin, A. G. and Tkachev, I. I.",
    title = "{Relativistic axions from collapsing Bose stars}",
    eprint = "1609.03611",
    archivePrefix = "arXiv",
    primaryClass = "astro-ph.CO",
    reportNumber = "INR-TH-2016-034",
    doi = "10.1103/PhysRevLett.118.011301",
    journal = "Phys. Rev. Lett.",
    volume = "118",
    number = "1",
    pages = "011301",
    year = "2017"
}

@article{deGiorgi:2024pjb,
    author = "de Giorgi, Arturo and Jaeckel, Joerg",
    title = "{Star shearing season: transient signals in wave-like dark matter experiments from black hole formation}",
    eprint = "2408.10296",
    archivePrefix = "arXiv",
    primaryClass = "hep-ph",
    reportNumber = "IFT-UAM/CSIC-24-122",
    doi = "10.1140/epjc/s10052-025-13872-0",
    journal = "Eur. Phys. J. C",
    volume = "85",
    number = "2",
    pages = "155",
    year = "2025"
}

@article{Eby:2024mhd,
    author = "Eby, Joshua and Takhistov, Volodymyr",
    title = "{Diffuse Axion Background}",
    eprint = "2402.00100",
    archivePrefix = "arXiv",
    primaryClass = "hep-ph",
    reportNumber = "KEK-QUP-2024-0001, KEK-TH-2596, KEK-Cosmo-0337, IPMU24-0003",
    month = "1",
    year = "2024"
}

@article{Eby:2021ece,
    author = "Eby, Joshua and Shirai, Satoshi and Stadnik, Yevgeny V. and Takhistov, Volodymyr",
    title = "{Probing relativistic axions from transient astrophysical sources}",
    eprint = "2106.14893",
    archivePrefix = "arXiv",
    primaryClass = "hep-ph",
    reportNumber = "IPMU21-0042",
    doi = "10.1016/j.physletb.2021.136858",
    journal = "Phys. Lett. B",
    volume = "825",
    pages = "136858",
    year = "2022"
}

@article{Eby:2016cnq,
    author = "Eby, Joshua and Leembruggen, Madelyn and Suranyi, Peter and Wijewardhana, L. C. R.",
    title = "{Collapse of Axion Stars}",
    eprint = "1608.06911",
    archivePrefix = "arXiv",
    primaryClass = "astro-ph.CO",
    reportNumber = "FERMILAB-PUB-16-678-T",
    doi = "10.1007/JHEP12(2016)066",
    journal = "JHEP",
    volume = "12",
    pages = "066",
    year = "2016"
}

@article{Hees:2018fpg,
    author = "Hees, Aur\'elien and Minazzoli, Olivier and Savalle, Etienne and Stadnik, Yevgeny V. and Wolf, Peter",
    title = "{Violation of the equivalence principle from light scalar dark matter}",
    eprint = "1807.04512",
    archivePrefix = "arXiv",
    primaryClass = "gr-qc",
    doi = "10.1103/PhysRevD.98.064051",
    journal = "Phys. Rev. D",
    volume = "98",
    number = "6",
    pages = "064051",
    year = "2018"
}

@article{KONOPLIV2011401,
title = {Mars high resolution gravity fields from MRO, Mars seasonal gravity, and other dynamical parameters},
journal = {Icarus},
volume = {211},
number = {1},
pages = {401-428},
year = {2011},
issn = {0019-1035},
doi = {https://doi.org/10.1016/j.icarus.2010.10.004},
url = {https://www.sciencedirect.com/science/article/pii/S0019103510003830},
author = {Alex S. Konopliv and Sami W. Asmar and William M. Folkner and Özgür Karatekin and Daniel C. Nunes and Suzanne E. Smrekar and Charles F. Yoder and Maria T. Zuber}
}

@inproceedings{Fischbach:1996eq,
    author = "Fischbach, Ephraim and Talmadge, Carrick",
    title = "{Ten years of the fifth force}",
    booktitle = "{31st Rencontres de Moriond: Dark Matter and Cosmology, Quantum Measurements and Experimental Gravitation}",
    eprint = "hep-ph/9606249",
    archivePrefix = "arXiv",
    reportNumber = "EFI-94-11A",
    pages = "443--451",
    year = "1996"
}

@article{Adelberger:2003zx,
    author = "Adelberger, E. G. and Heckel, Blayne R. and Nelson, A. E.",
    title = "{Tests of the gravitational inverse square law}",
    eprint = "hep-ph/0307284",
    archivePrefix = "arXiv",
    doi = "10.1146/annurev.nucl.53.041002.110503",
    journal = "Ann. Rev. Nucl. Part. Sci.",
    volume = "53",
    pages = "77--121",
    year = "2003"
}

@article{Berge:2017ovy,
    author = "Berg\'e, Joel and Brax, Philippe and M\'etris, Gilles and Pernot-Borr\`as, Martin and Touboul, Pierre and Uzan, Jean-Philippe",
    title = "{MICROSCOPE Mission: First Constraints on the Violation of the Weak Equivalence Principle by a Light Scalar Dilaton}",
    eprint = "1712.00483",
    archivePrefix = "arXiv",
    primaryClass = "gr-qc",
    doi = "10.1103/PhysRevLett.120.141101",
    journal = "Phys. Rev. Lett.",
    volume = "120",
    number = "14",
    pages = "141101",
    year = "2018"
}

@article{Dailey:2020sxa,
    author = "Dailey, Conner and Bradley, Colin and Jackson Kimball, Derek F. and Sulai, Ibrahim A. and Pustelny, Szymon and Wickenbrock, Arne and Derevianko, Andrei",
    title = "{Quantum sensor networks as exotic field telescopes for multi-messenger astronomy}",
    eprint = "2002.04352",
    archivePrefix = "arXiv",
    primaryClass = "astro-ph.IM",
    doi = "10.1038/s41550-020-01242-7",
    journal = "Nature Astron.",
    volume = "5",
    number = "2",
    pages = "150--158",
    year = "2021"
}

@article{Stadnik:2021qyf,
    author = "Stadnik, Yevgeny V.",
    title = "{Highlighting the back-action contribution of matter to quantum sensor network performance in multi-messenger astronomy}",
    eprint = "2111.14351",
    archivePrefix = "arXiv",
    primaryClass = "gr-qc",
    doi = "10.1038/s41550-024-02245-4",
    journal = "Nature Astron.",
    volume = "8",
    number = "4",
    pages = "434--438",
    year = "2024"
}

@article{Nakanishi_2015,
   title={Three-dimensional distribution of the ISM in the Milky Way galaxy. III. The total neutral gas disk},
   volume={68},
   ISSN={0004-6264},
   url={http://dx.doi.org/10.1093/pasj/psv108},
   DOI={10.1093/pasj/psv108},
   number={1},
   journal={Publications of the Astronomical Society of Japan},
   publisher={Oxford University Press (OUP)},
   author={Nakanishi, Hiroyuki and Sofue, Yoshiaki},
   year={2015},
   month=nov }

@article{schechter,
       author = {{Schechter}, P.},
        title = "{An analytic expression for the luminosity function for galaxies.}",
      journal = {APJ},
         year = 1976,
        month = jan,
       volume = {203},
        pages = {297-306},
          doi = {10.1086/154079},
       adsurl = {https://ui.adsabs.harvard.edu/abs/1976ApJ...203..297S}
}

@article{Arakawa:2023gyq,
    author = "Arakawa, Jason and Eby, Joshua and Safronova, Marianna S. and Takhistov, Volodymyr and Zaheer, Muhammad H.",
    title = "{Detection of bosenovae with quantum sensors on Earth and in space}",
    eprint = "2306.16468",
    archivePrefix = "arXiv",
    primaryClass = "hep-ph",
    reportNumber = "IPMU23-0024, KEK-QUP-2023-0014, KEK-TH-2534, KEK-Cosmo-0316",
    doi = "10.1103/PhysRevD.110.075007",
    journal = "Phys. Rev. D",
    volume = "110",
    number = "7",
    pages = "075007",
    year = "2024"
}

@article{Arakawa:2024lqr,
    author = "Arakawa, Jason and Zaheer, Muhammad H. and Eby, Joshua and Takhistov, Volodymyr and Safronova, Marianna S.",
    title = "{Bosenovae with quadratically-coupled scalars in quantum sensing experiments}",
    eprint = "2402.06736",
    archivePrefix = "arXiv",
    primaryClass = "hep-ph",
    reportNumber = "KEK-QUP-2024-0002, KEK-TH-2599, KEK-Cosmo-0338, IPMU24-0004",
    doi = "10.1007/JHEP08(2024)222",
    journal = "JHEP",
    volume = "08",
    pages = "222",
    year = "2024"
}

@article{Scherer:2002tk,
    author = "Scherer, Stefan",
    editor = "Negele, John W. and Vogt, E. W.",
    title = "{Introduction to chiral perturbation theory}",
    eprint = "hep-ph/0210398",
    archivePrefix = "arXiv",
    reportNumber = "MKPH-T-02-09",
    journal = "Adv. Nucl. Phys.",
    volume = "27",
    pages = "277",
    year = "2003"
}

@article{Arakawa:2025hcn,
    author = "Arakawa, Jason and Zaheer, Muhammad H. and Takhistov, Volodymyr and Safronova, Marianna S. and Eby, Joshua and Cheung, Charles",
    title = "{Multimessenger astronomy beyond the Standard Model: New window from quantum sensors}",
    eprint = "2502.08716",
    archivePrefix = "arXiv",
    primaryClass = "hep-ph",
    reportNumber = "KEK-QUP-2025-0001, KEK-TH-2688, KEK-Cosmo-0371",
    doi = "10.1088/1475-7516/2026/02/026",
    journal = "JCAP",
    volume = "02",
    pages = "026",
    year = "2026"
}

@article{Khamis:2024oqa,
    author = "Khamis, Sami S. and others",
    title = "{Multimessenger Search for Exotic Field Emission with a Global Magnetometer Network}",
    eprint = "2407.13919",
    archivePrefix = "arXiv",
    primaryClass = "astro-ph.HE",
    doi = "10.1103/fsc8-6sl5",
    journal = "Phys. Rev. X",
    volume = "15",
    number = "3",
    pages = "031048",
    year = "2025"
}

@article{deGiorgi:2026lez,
    author = "de Giorgi, Arturo and del Castillo, Yeray Garcia and Jaeckel, Joerg",
    title = "{Scalar tsunamis from black hole formation}",
    eprint = "2602.06112",
    archivePrefix = "arXiv",
    primaryClass = "hep-ph",
    reportNumber = "IPPP/26/11",
    doi = "10.1088/1475-7516/2026/09/127",
    journal = "JCAP",
    volume = "09",
    pages = "127",
    year = "2026"
}

@article{Panelli_2020,
   title={Applying the matched-filter technique to the search for dark matter transients with networks of quantum sensors},
   volume={7},
   ISSN={2196-0763},
   url={http://dx.doi.org/10.1140/epjqt/s40507-020-00081-9},
   DOI={10.1140/epjqt/s40507-020-00081-9},
   number={1},
   journal={EPJ Quantum Technology},
   publisher={Springer Science and Business Media LLC},
   author={Panelli, Guglielmo and Roberts, Benjamin M. and Derevianko, Andrei},
   year={2020},
   month=Feb
   }

@article{LIGOScientific:2016aoc,
    author = "Abbott, B. P. and others",
    collaboration = "LIGO Scientific, Virgo",
    title = "{Observation of Gravitational Waves from a Binary Black Hole Merger}",
    eprint = "1602.03837",
    archivePrefix = "arXiv",
    primaryClass = "gr-qc",
    reportNumber = "LIGO-P150914",
    doi = "10.1103/PhysRevLett.116.061102",
    journal = "Phys. Rev. Lett.",
    volume = "116",
    number = "6",
    pages = "061102",
    year = "2016"
}

@article{Romano:2016dpx,
    author = "Romano, Joseph D. and Cornish, Neil J.",
    title = "{Detection methods for stochastic gravitational-wave backgrounds: a unified treatment}",
    eprint = "1608.06889",
    archivePrefix = "arXiv",
    primaryClass = "gr-qc",
    doi = "10.1007/s41114-017-0004-1",
    journal = "Living Rev. Rel.",
    volume = "20",
    number = "1",
    pages = "2",
    year = "2017"
}

@misc{AxionLimits,
  author       = {Ciaran O'Hare},
  title        = {cajohare/AxionLimits: AxionLimits},
  month        = jul,
  year         = 2020,
  publisher    = {Zenodo},
  version      = {v1.0},
  doi          = {10.5281/zenodo.3932430},
  howpublished = {\url{https://cajohare.github.io/AxionLimits/}}
}

@article{Gron:1995rn,
    author = "Gron, Oyvind and Soleng, Harald H.",
    title = "{Experimental limits to the density of dark matter in the solar system}",
    eprint = "astro-ph/9507051",
    archivePrefix = "arXiv",
    reportNumber = "CERN-TH-7500-94-REV, CERN-TH-7500-94",
    doi = "10.1086/176669",
    journal = "Astrophys. J.",
    volume = "456",
    pages = "445--448",
    year = "1996"
}

@article{Anderson:1995dw,
    author = "Anderson, John D. and Lau, Eunice L. and Krisher, Timothy P. and Dicus, Duane A. and Rosenbaum, Doris C. and Teplitz, Vigdor L.",
    title = "{Improved bounds on nonluminous matter in solar orbit}",
    eprint = "hep-ph/9503368",
    archivePrefix = "arXiv",
    reportNumber = "CPP-93-18",
    doi = "10.1086/176017",
    journal = "Astrophys. J.",
    volume = "448",
    pages = "885",
    year = "1995"
}

@article{Khriplovich:2006sq,
    author = "Khriplovich, Iosif B. and Pitjeva, E. V.",
    editor = "Kleinert, Hagen and Jantzen, Robert T. and Ruffini, Remo",
    title = "{Upper limits on density of dark matter in solar system}",
    eprint = "astro-ph/0601422",
    archivePrefix = "arXiv",
    doi = "10.1142/S0218271806008462",
    journal = "Int. J. Mod. Phys. D",
    volume = "15",
    pages = "615--618",
    year = "2006"
}

@article{Frere:2007pi,
    author = "Frere, J. -M. and Ling, Fu-Sin and Vertongen, G.",
    title = "{Bound on the Dark Matter Density in the Solar System from Planetary Motions}",
    eprint = "astro-ph/0701542",
    archivePrefix = "arXiv",
    reportNumber = "ULB-TH-07-04",
    doi = "10.1103/PhysRevD.77.083005",
    journal = "Phys. Rev. D",
    volume = "77",
    pages = "083005",
    year = "2008"
}

@article{Pitjev:2013sfa,
    author = "Pitjev, N. P. and Pitjeva, E. V.",
    title = "{Constraints on dark matter in the solar system}",
    eprint = "1306.5534",
    archivePrefix = "arXiv",
    primaryClass = "astro-ph.EP",
    doi = "10.1134/S1063773713020060",
    journal = "Astron. Lett.",
    volume = "39",
    pages = "141--149",
    year = "2013"
}

@article{Tsai:2022jnv,
    author = "Tsai, Yu-Dai and Eby, Joshua and Arakawa, Jason and Farnocchia, Davide and Safronova, Marianna S.",
    title = "{OSIRIS-REx constraints on local dark matter and cosmic neutrino profiles}",
    eprint = "2210.03749",
    archivePrefix = "arXiv",
    primaryClass = "hep-ph",
    reportNumber = "UCI-HEP-TR-2022-11, FERMILAB-PUB-22-753-T-V",
    doi = "10.1088/1475-7516/2024/02/029",
    journal = "JCAP",
    volume = "02",
    pages = "029",
    year = "2024"
}

@misc{githubGitHubAutomerisioWebPlotDigitizer,
	author = {},
	title = {{G}it{H}ub - automeris-io/{W}eb{P}lot{D}igitizer: {C}omputer vision assisted tool to extract numerical data from plot images. --- github.com},
	howpublished = {\url{https://github.com/automeris-io/{W}eb{P}lot{D}igitizer}},
	year = {},
	note = {[Accessed 21-07-2026]},
}

\newpage

\end{document}